\documentclass[journal]{IEEEtran}
\usepackage{cite,graphicx,amsmath,psfrag,bm}

\usepackage[usenames,dvipsnames]{xcolor}
\usepackage{mathabx}
\usepackage{amsbsy}
\usepackage{soul}
\usepackage{caption}
\usepackage{subcaption}
\usepackage{setspace}
\usepackage{rotating}
\usepackage[electronic]{ifsym}
\usepackage{moresize}
\usepackage{mathtools}
\usepackage{tikz}
\usetikzlibrary{matrix}
\usepackage[mode=errorstop]{pstool}
\usepackage{psfrag}
\usepackage{epstopdf}
\usepackage[normalem]{ulem}

\newcommand{\ve}[1]{\boldsymbol{#1}}
\newcommand{\te}[1]{\overline{\overline{#1}}}

\begin{document}

\title{Metasurface Modeling by a Thin Slab}

\author{Mojtaba Dehmollaian,~\IEEEmembership{Senior Member}, IEEE, Yousef~Vahabzadeh, Karim~Achouri, and~Christophe~Caloz,~\IEEEmembership{Fellow,~IEEE}

}


\maketitle

\begin{abstract}
We investigate the possibility to model a metasurface, defined as a zero-thickness sheet of surface polarization currents, by a thin slab, characterized by a subwavelength thickness and usual voluminal medium parameters. First, we elaborate a general equivalence relation between the metasurface and the slab in terms of average electromagnetic fields. Then, we derive exact relations between the metasurface and slab susceptibilities and validate them by full-wave simulations. Finally, we discuss the simple and insightful Average Field Approximation (AFA) formula, illustrate its inappropriate for strong metasurface field transformations, and establish its range of validity. All of these developments are restricted to the simplest case of a uniform isotropic metasurface under normal plane wave incidence. We conclude from the complexity of the equivalence for this case, that a metasurface is generally best modeled in terms of Generalized Sheet Transition Conditions (GSTCs). 
\end{abstract}

\begin{IEEEkeywords}
Metasurface, Generalized Sheet Transition Conditions (GSTCs), metamasurface modeling, thin slab.
\end{IEEEkeywords}

\IEEEpeerreviewmaketitle

\section{Introduction}
Metasurfaces~\cite{Achouri_EMTA_2020,Holloway_ms_applic_review_2012, Glybovski_metasurfaces_2016, Achouri_EPJAM_01_2016} have recently attracted massive attention due to their unique properties~\cite{Alu_light_pol_manipul_plasm_ms_2011, Kildishev_planar_photonic_ms_2013, Grbic_bianiso_ms_2014, Zheng_2015_ms_hologram_80_effi} and numerous applications, which include, for instance, dispersion phase compensation~\cite{Aieta_achromatic_MS_2019}, wide-angle Fourier transformation~\cite{Liu_FT_ms_2018}, spatial mixing~\cite{Won_mixer_ms_2018}, ultra-violet vacuum light generation~\cite{Semmlinger_vacuum_UV_ms_2018} and real-time spacetime processing~\cite{Chamanara_TAP_04_2019}.

Practical metasurfaces have a deeply subwavelength thickness, i.e., a physical thickness, $d$, that is much smaller than the wavelength of the incident wave, $\lambda$, namely $d\ll\lambda$. Therefore, they do not support any Fabry-Perot type resonances, which would require $d>\lambda/4$, and really behave as polarization current \emph{sheets} with zero thickness\footnote{In microwave engineering language, one would say that the structure is \emph{lumped} in terms of thickness, while being largely distributed in its plane~\cite{Pozar_ME_2011}.}~\cite{Achouri_EMTA_2020,Holloway_ms_applic_review_2012,Glybovski_metasurfaces_2016}. They are therefore perfectly modeled by Generalized Sheet Transition Conditions (GSTCs)~\cite{Achouri_EMTA_2020,Holloway_ms_applic_review_2012,Glybovski_metasurfaces_2016,Keuster, idemen, Kuester_averag_trans_metafilm_2003, Achouri_general_ms_synthes_2015}\footnote{Directly full-wave simulating a metasurface with its metaparticles is naturally possible, but this would be a highly unproductive approach because it is extremely lengthy or/and unstable, given the super-dense meshing required to account for the subwavelength thickness of the metasurface, and because it provides little insight into the fundamental operation and limits of the specific structure that is simulated, which makes the design of sophisticated metasurfaces quasi-impossible. The most efficient metasurface synthesis technique existing to date is the two-step procedure, which consists in first determining the homogeneous susceptibility tensor functions realizing specified transformation, and second discretizing these functions and determining the proper particle shapes via scattering mapping~\cite{Achouri_EMTA_2020, Achouri_NP_06_2018}.}.

At the development time of the GSTCs for metasurfaces -- and this is in fact still the case at the time of this writing! -- no commercial software was able to simulate bianisotropic polarization sheets, and a benchmarking model was therefore highly needed. For this reason, and having no better choice, we resorted to the model of the metasurface by a subwavenengthly thin slab, where the local effect of the metasurface sheet was diluted across the slab thickness~\cite{Liu_homogenized_MS_2018}, i.e., $\chi_\text{v}=\chi/d$, with $\chi_\text{v}$ being the voluminal susceptibility of the slab and $\chi_\text{v}=\epsilon_\text{r}-1$ ($\epsilon_\text{r}$: relative permittivity) in the case of simple dielectric slab~\cite{Yousef_FDFD_2016,Yousef_FDTD_2018,Yousef_Comp_analy_ms_2018}. The ``diluted-slab'' approached clearly makes sense for metasurface transformations of moderate strength, and indeed allowed some level of benchmarking in the aforementioned references, but also clear appears to be questionable in the case of strong metasurface transformations, such as for instance in the cases of a  metasurface absorber or gyrator. A detailed investigation of this issue has been completely missing in the literature on metasurfaces to date. This paper fills up this gap by presenting an in-depth analysis of the problem.

The paper is organized as follows. Section~\ref{sec:sta_prob} states the problem of equivalence between a metasurface and a thin slab, and introduces assumptions holding throughout the document. Section~\ref{sec:rec_met_synth} recalls the basic GSTC equations and writes them explicitly under the assumptions of the previous section. Section~\ref{sec:equiv} elaborates a general condition for metasurface-slab equivalence in terms of corresponding average fields. Section~\ref{sec:exp_ex_rel} derives explicit exact relations between the metasurface and slab susceptibilities and validates these relations by full-wave simulations. Section~\ref{sec:equiv} examines and discusses the ``diluted-slab'' approximation, called here the Average Field Approximation (AFA) and establishes its range of validity. Finally, Sec.~\ref{sec:equiv} concludes the paper.

\section{Statement of the Problem}\label{sec:sta_prob}

\begin{figure}[!ht]
	\centering
	\begin{subfigure}{1\columnwidth}
		\centering
		\includegraphics[width=0.6\columnwidth]{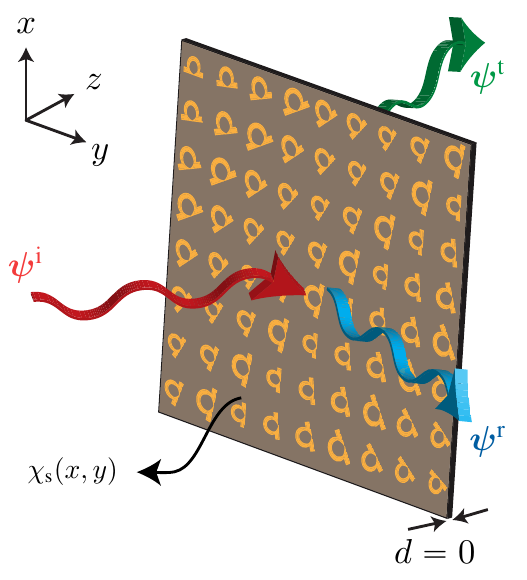}{
		}\caption{}\label{Fig:Metasurface_GSTC}
	\end{subfigure}
	\begin{subfigure}{1\columnwidth}
		\centering
		\includegraphics[width=0.6\columnwidth]{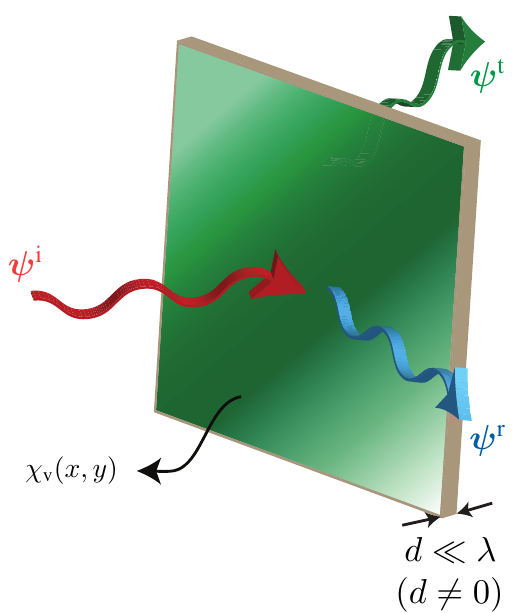}{
			\psfrag{i}[][][1.]{\textcolor[rgb]{0,0.6,0.275}{$\ve{\psi}^\textrm{t}$}}
			\psfrag{j}[][][1.]{$\te{\chi}(\ve{x,y})$}
			\psfrag{y}[][][1.]{$(d\neq0)$}
			\psfrag{k}[][][1.]{$d\ll \lambda$}
			\psfrag{g}[][][1.]{\textcolor[rgb]{0.953,0.23,0.23}{$\ve{\psi}^\textrm{i}$}}
			\psfrag{h}[][][1.]{\textcolor[rgb]{0,0.365,0.635}{$\ve{\psi}^\textrm{r}$}}
			\psfrag{x}[r][r][0.8]{$\chi_\text{v}(x,y)$}
		} \caption{}\label{Fig:Metasurface_slab}
	\end{subfigure}
	\caption{Modeling of metasurface by a subwavelength slab. (a)~Metasurface, with surface susceptibility $\chi(x,y)$. (b)~Slab, with subwavelength thickness $d$ ($d\ll\lambda$) and $z$-uniform (or constant) voluminal susceptibility, $\chi_\text{v}(x,y)$, with $\chi_\text{v}\neq\chi_\text{v}(z)$, intended to model~(a).} \label{Fig:Metasurface}
\end{figure}

The problem to solve is depicted in Fig.~\ref{Fig:Metasurface}. It consists in modeling the metasurface represented in Fig.~\ref{Fig:Metasurface_GSTC} by the thin slab shown in Fig.~\ref{Fig:Metasurface_slab}. The metasurface is defined as a zero-thickness sheet formed by a two-dimensional array of subwavelength scattering particles and modeled by an homogenized surface (possibly tensorial) susceptibility function, $\chi_\text{s}(x,y)$, which we simply denote $\chi(x,y)$ in the sequel of the paper; it transforms an incident wave $\boldsymbol{\psi}^\textrm{i}$ into a reflected wave $\boldsymbol{\psi}^\textrm{r}$ and a transmitted wave $\boldsymbol{\psi}^\textrm{t}$. The slab has a subwavelength thickness $d$ ($d\ll\lambda$), to ensure the absence of Fabry-Perot resonances, and is modeled by the susceptibility, $\chi_\text{v}(x,y)$.

\begin{figure}[h]
	\centering
	\includegraphics[width=1\columnwidth]{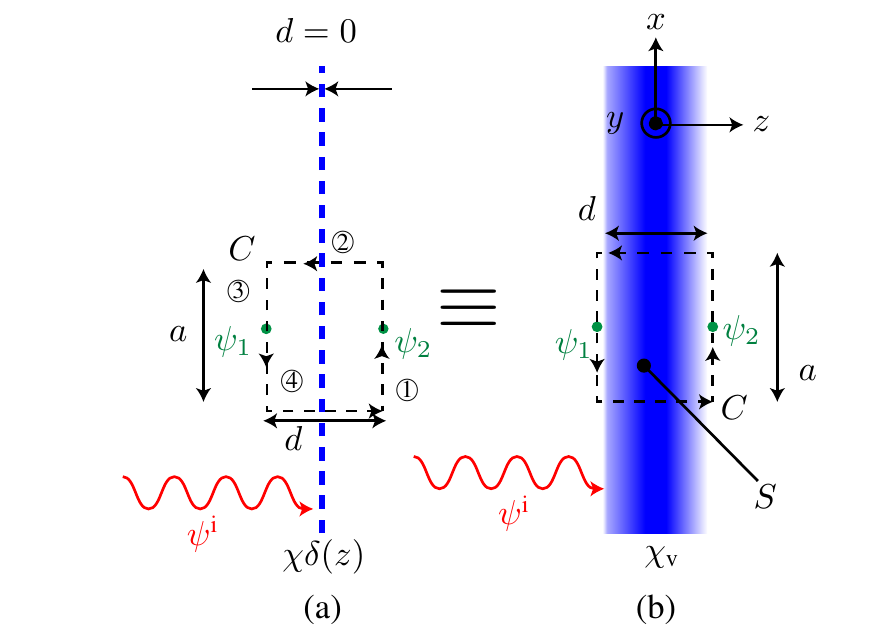}{
    }
	\caption{Mathematical setup used to investigate the conditions for electromagnetic equivalence between a metasurface and a subwavelength slab, under the assumptions enunciated in Sec.~\ref{sec:sta_prob}. (a)~Metasurface, with surface susceptibility $\chi$, corresponding to Fig.~\ref{Fig:Metasurface_GSTC}. (b)~Approximating slab, of subwavelength thickness $d$, with volume susceptibility $\chi_\text{v}$, corresponding to Fig.~\ref{Fig:Metasurface_slab}.} \label{Fig:diluted_ms_approx}
\end{figure}

For simplicity, we make the following assumptions:
\begin{enumerate}
	\item The metasurface structure is uniform along the $y$-direction.
	\item The excitation is plane wave that normally impinges on the metasurface with polarization $(E_y,H_x)$ and $E_z=E_x=H_y=H_z=0$.
	\item As a consequence of the previous two assumptions, the problem reduces to a two-dimensional problem, with $\partial/\partial{y}=0$ and scattering in the $(z,x)$ plane, as shown in Fig.~\ref{Fig:diluted_ms_approx}.
	\item The metasurface is homoisotropic, i.e., has $\chi_\textrm{ee},\chi_\textrm{mm}\neq{0}$ with both quantities being scalar, and $\chi_\textrm{em}=\chi_\textrm{me}=0$.
	\item The metasurface is initially also uniform along the $x$-direction, i.e., $\chi\neq\chi(x,y)$, which implies purely normal scattering.
\end{enumerate}

\section{GSTC Modeling of a Metasurface}\label{sec:rec_met_synth}
The time-harmonic GSTC equations for the metasurface in Fig.~\ref{Fig:Metasurface}a, assuming for simplicity zero normal surface susceptibilities, are~\cite{Achouri_EMTA_2020,Mojtaba_GSTC_TBC_2019,Achouri_NP_06_2018,Achouri_general_ms_synthes_2015}
\begin{subequations}\label{GSTC}
\begin{align}\label{GSTC1}
  \begin{pmatrix}
    -\Delta H_y \\
    \Delta H_x \\
  \end{pmatrix}
=j\omega\epsilon_0 &
                        \begin{pmatrix}
                          \chi _{\textrm{ee}}^{xx} & \chi _{\textrm{ee}}^{xy} \\
                          \chi _{\textrm{ee}}^{yx} & \chi _{\textrm{ee}}^{yy} \\
                        \end{pmatrix}
                        \begin{pmatrix}
                          E_{x,\textrm{av}} \\
                          E_{y,\textrm{av}} \\
                        \end{pmatrix}
                      \\ \notag
                      &+j\omega\sqrt{\epsilon_0 \mu_0}
                                                           \begin{pmatrix}
                                                             \chi_{\textrm{em}}^{xx} & \chi_{\textrm{em}}^{xy} \\
                                                             \chi_{\textrm{em}}^{yx} & \chi_{\textrm{em}}^{yy} \\
                                                           \end{pmatrix}
                                                           \begin{pmatrix}
                                                             H_{x,\textrm{av}} \\
                                                             H_{y,\textrm{av}} \\
                                                           \end{pmatrix},
  \end{align}
  \begin{align}\label{GSTC2}
  \begin{pmatrix}
    \Delta E_y \\
    -\Delta E_x \\
  \end{pmatrix}
 =j\omega\mu_0 &
                        \begin{pmatrix}
                          \chi _{\textrm{mm}}^{xx} & \chi _{\textrm{mm}}^{xy} \\
                          \chi _{\textrm{mm}}^{yx} & \chi _{\textrm{mm}}^{yy} \\
                        \end{pmatrix}
                        \begin{pmatrix}
                          H_{x,\textrm{av}} \\
                          H_{y,\textrm{av}} \\
                        \end{pmatrix}
                      \\\notag
                      &+j\omega\sqrt{\epsilon_0 \mu_0}
                                                           \begin{pmatrix}
                                                             \chi_{\textrm{me}}^{xx} & \chi _{\textrm{me}}^{xy} \\
                                                             \chi_{\textrm{me}}^{yx} & \chi _{\textrm{me}}^{yy} \\
                                                           \end{pmatrix}
                                                           \begin{pmatrix}
                                                             E_{x,\textrm{av}} \\
                                                             E_{y,\textrm{av}}
                                                           \end{pmatrix},
\end{align}
\end{subequations}
where $\Delta\psi=\psi^\text{t}-\left(\psi^\text{i}+\psi^\text{r}\right)$ and $\psi_\text{av}=[\left(\psi^\text{i}+\psi^\text{r}\right)+\psi^\text{t}]/2$, with $\psi$ representing any component of the tangential electric field ($E_x,E_y$) or magnetic field ($H_x,H_y$), with i, r, t, e and m standing for incident, reflected, transmitted, electric and magnetic, respectively, and where $\omega$ is the angular frequency, $\epsilon_0$ is the free-space permittivity and $\mu_0$ is the free-space permeability.

Under the assumptions enunciated in Sec.~\ref{sec:sta_prob}, the equations~\eqref{GSTC} reduce to
\begin{subequations}\label{eq:2D_GSTC}
\begin{align}
  \Delta H_x^\text{m}&=j\omega\epsilon_0\chi_\text{ee}^{yy}E_{y,\text{av}}^\text{m},\label{eq:2D_GSTC_A}  \\
  \Delta E_y^\text{m}&=j\omega\mu_0\chi_\text{mm}^{xx}H_{x,\text{av}}^\text{m},
\end{align}
\end{subequations}
with
\begin{subequations}\label{eq:EH_difav}
\begin{equation}
\Delta H_x^\text{m}=H_x^\text{t}-\left(H_x^\text{i}+H_x^\text{r}\right),
\end{equation}
\begin{equation}\label{eq:E_difava}
E_{y,\text{av}}^\text{m}=\frac{\left(E_y^\text{i}+E_y^\text{r}\right)+E_y^\text{t}}{2},
\end{equation}
\begin{equation}
\Delta E_y^\text{m}=E_y^\text{t}-\left(E_y^\text{i}+E_y^\text{r}\right),
\end{equation}
\begin{equation}\label{eq:H_difava}
H_{x,\text{av}}^\text{m}=\frac{\left(H_x^\text{i}+H_x^\text{r}\right)+H_x^\text{t}}{2},
\end{equation}
\end{subequations}
where $\chi_\text{ee}^{yy}$ and $\chi_\text{mm}^{xx}$ are constant, and where we have introduced the superscript `m', standing for metasurface, for later distinction with the slab fields.

\section{Metasurface -- Slab Equivalence Relations}\label{sec:equiv}
Figure~\ref{Fig:diluted_ms_approx} shows the mathematical setup that we shall use to investigate the conditions for electromagnetic equivalence between a metasurface and a subwavelength slab, under the assumptions that were enunciated in Sec.~\ref{sec:sta_prob}. Figure~\ref{Fig:diluted_ms_approx}a corresponds to the metasurface, with surface susceptibility $\chi$, while Fig.~\ref{Fig:diluted_ms_approx}b corresponds to the approximating slab, with thickness $d$ and with volume susceptibility $\chi_\text{v}$. Both cases involve a rectangular integration surface $S$, with width $d$ corresponding to the width of the slab and delimited by the contour~$C$.

The fields in Figs.~\ref{Fig:diluted_ms_approx}a and~\ref{Fig:diluted_ms_approx}b must both satisfy the Maxwell equations, and in particular the Maxwell-Amp\`{e}re law, whose integral form reads
\begin{equation}\label{eq:stokes}
  \oint\limits_{C} \mathbf{H}\cdot\mathrm{d}\mathbf{l}
  =j\omega\iint\limits_{S}\mathbf{D}\cdot\mathrm{d}\mathbf{S}+\iint\limits_{S}\mathbf{J}\cdot\mathrm{d}\mathbf{S},
\end{equation}
where the boldface font indicates vector quantities. We shall next assume that there is no (external) source ($\ve{J}=0$) on the metasurface, as is most often the case in metasurface applications, and hence also in the slab intended to approximate it, which eliminates the second integral in the right-hand side of~\eqref{eq:stokes}.

In a (voluminal) dielectric medium, such as the medium forming the slab in Fig.~\ref{Fig:diluted_ms_approx}b, the displacement field, which accounts for the response of the medium, is related to the electric field via the usual formula
\begin{subequations}
\begin{equation}\label{eq:DEv}
\mathbf{D}=\epsilon\mathbf{E}=\epsilon_0(1+\chi_\text{v})\mathbf{E},
\end{equation}
which simplifies here to
\begin{equation}
D_y=\epsilon_0(1+\chi_\text{v})E_y.
\end{equation}
\end{subequations}
In contrast, in a two-dimensional medium, such as the (zero-thickness) metasurface in Fig.~\ref{Fig:diluted_ms_approx}a, this relation takes the form
\begin{subequations}
\begin{equation}
\mathbf{D}=\epsilon_0\left[1+\chi\delta(z)\right]\mathbf{E},
\end{equation}
which simplifies here to
\begin{equation}\label{eq:Dms}
D_y=\epsilon_0\left[1+\chi\delta(z)\right]E_y,
\end{equation}
\end{subequations}
which incidentally reveals that $[\chi]=\text{m}$, since $[\chi \delta(z)]=[\chi_\text{v}]=1$ and since $[\delta(z)]=1/[z]=1/\text{m}$ according to the Dirac distribution definition $\int_{-\infty}^{+\infty}\delta(z)dz=1$ (unitless).

For the metasurface in Fig.~\ref{Fig:diluted_ms_approx}a, renaming $\chi$ in~\eqref{eq:Dms} $\chi_\text{ee}^{yy}$ for consistency with~\eqref{eq:2D_GSTC_A}, and substituting the resulting expression into the right-hand side of~\eqref{eq:stokes} (with $\ve{J}=0$), leads to
\begin{equation}\label{eq:LHS_stokes_simplified}
\begin{split}
\oint\limits_{C}\mathbf{H}\cdot\mathrm{d}\mathbf{l}
&=j\omega\epsilon_0\int_{-a/2}^{a/2}\chi_\text{ee}^{yy}E_{y,\text{av}}^\text{m}{\rm d}x \\
&\qquad\qquad+j\omega\epsilon_0\int_{-a/2}^{a/2}\int_{-d/2}^{d/2} E_y^{(0)}(z){\rm d}z{\rm d}x.
\end{split}
\end{equation}
In this relation, the result  $\int_{-d/2}^{+d/2}\delta(z)dz=1$ was used in the first equation of the right-hand side term, and the second integral of the right-hand side term represents the free-space contribution of the flux of $\mathbf{D}$ (not related to the metasurface per se) across the surface $S$ of the integration rectangle, with the free-space field denoted by the superscript $(0)$. Given the slab subwavelength thickness assumption, $d\ll\lambda_0$, the free-space field $E_y^{(0)}$ may be expressed as the average of the field within $S$, which is here the field on the metasurface, i.e.,
\begin{equation}\label{eq:approx_Ey_freee_space}
  \int_{-d/2}^{d/2} E_y^{(0)}\mathrm{d}z
  =E_{y,\text{av}}\int_{-d/2}^{d/2}\mathrm{d}z
  =E_{y,\text{av}}^\text{m}d,
\end{equation}
which reduces~\eqref{eq:LHS_stokes_simplified} to
\begin{equation}\label{eq:MS_Amp}
\oint\limits_{C}\mathbf{H}\cdot\mathrm{d}\mathbf{l}
=j\omega\epsilon_0\int_{-a/2}^{a/2}\left(\chi_\text{ee}^{yy}+d\right)E_{y,\text{av}}^\text{m}{\rm d}x.
\end{equation}

For the slab in Fig.~\ref{Fig:diluted_ms_approx}b, Eq.~\eqref{eq:stokes} (with $\ve{J}=0$) becomes with~\eqref{eq:DEv}
\begin{align}\label{eq:slab_Amp}
\oint\limits_{C} \mathbf{H}\cdot\mathrm{d}\mathbf{l}
&=j\omega\epsilon_0\left(1+\chi_\text{v,ee}\right)\int_{-a/2}^{a/2}\int_{-d/2}^{d/2}E_y^\text{s}(z)\mathrm{d} z\mathrm{d} x,
\end{align}
where the  subscript `ee' has been added for consistency with~\eqref{eq:MS_Amp} and where the superscript `s' has been introduced for specific reference to the slab.

Equating the right-hand sides of~\eqref{eq:MS_Amp} and~\eqref{eq:slab_Amp} to ensure equivalence between the metasurface and the slab, and dividing both sides of the resulting equation by~$j\omega\epsilon_0$, yields
\begin{equation}
\begin{split}
\int_{-a/2}^{a/2}&\left(\chi_\text{ee}^{yy}+d\right)E_{y,\text{av}}^\text{m}{\rm d}x \\
&\qquad=\left(1+\chi_\text{v,ee}\right)\int_{-a/2}^{a/2}\int_{-d/2}^{d/2}E_y^\text{s}(z)\mathrm{d} z\mathrm{d} x.
\end{split}
\end{equation}

Let us now move the right-hand side of this relation to the left side, and factor out the integral over $x$. This gives
\begin{equation}\label{eq:stokes_GSTC_equ}
\begin{split}
  \int_{-a/2}^{a/2}
  &\Biggl[\left(\chi_\text{ee}^{yy}+d\right)E_{y,\text{av}}^\text{m} \\ &\qquad-\left(1+\chi_\text{v,ee}\right)\int_{-d/2}^{d/2}E_y^\text{s}(z)\mathrm{d} z\Biggr]\mathrm{d} x=0.
\end{split}
\end{equation}
The integrand in the left-hand side of Eq.~{\eqref{eq:stokes_GSTC_equ}} must be an odd function of $x$ for the integral to vanish according to the right-hand side. Moreover, since the choice of the coordinate center along the $x$ axis is arbitrary, given the $x$-uniformity assumption, it even needs to be odd for any choice of the origin along the $x$ axis. Therefore, it can only be zero, so that
\begin{subequations}\label{eq:equiv_rel0}
\begin{equation}\label{eq:stokes_GSTC_solution0}
  \left(\chi_\text{ee}^{yy}+d\right)E_{y,\text{av}}^\text{m}=\left(1+\chi_\text{v,ee}\right)\int_{-d/2}^{d/2}E_y^\text{s}(z)\mathrm{d} z.
\end{equation}

A similar procedure, using now the Maxwell-Faraday law, leads to the dual relation
\begin{equation}\label{eq:stokes_GSTC_mgnetic0}
  \left(\chi_\text{mm}^{xx}+d\right)H_{x,\text{av}}^\text{m}=\left(1+\chi_\text{v,mm}\right)\int_{-d/2}^{d/2}H_x^\text{s}(z)\mathrm{d} z.
\end{equation}
\end{subequations}

Defining the field averages across the slab as
\begin{subequations}
	\begin{equation}\label{eq:Eyavs}
	E_{\text{av},y}^\text{s}=\frac{1}{d}\int_{-d/2}^{d/2}E_y^\text{s}(z)\mathrm{d} z,
	\end{equation}
	\begin{equation}\label{eq:Hxavs}
	H_{\text{av},x}^\text{s}=\frac{1}{d}\int_{-d/2}^{d/2}H_x^\text{s}(z)\mathrm{d} z,
	\end{equation}
\end{subequations}
the relations~\eqref{eq:equiv_rel0} can be reformulated in the more explicit form
\begin{subequations}\label{eq:equiv_rel}
	\begin{equation}\label{eq:stokes_GSTC_solution}
	\left(\chi_\text{ee}^{yy}+d\right)E_{y,\text{av}}^\text{m}=d\left(1+\chi_\text{v,ee}\right)E_{\text{y,av}}^\text{s},
	\end{equation}
	\begin{equation}\label{eq:stokes_GSTC_mgnetic}
	\left(\chi_\text{mm}^{xx}+d\right)H_{x,\text{av}}^\text{m}=d\left(1+\chi_\text{v,mm}\right)H_{\text{x,av}}^\text{s},
	\end{equation}
\end{subequations}
whose resolution for the volume susceptibilities yields
\begin{subequations}\label{eq:av_ratio_chi}    
	\begin{equation}\label{eq:av_ratio_chi_e}
	\chi_\text{v,ee}
	=\left(1+\frac{\chi_\text{ee}^{yy}}{d}\right)\frac{E_{y,\text{av}}^\text{m}}{E_{y,\text{av}}^\text{s}}-1,
	\end{equation}
	\begin{equation}\label{eq:av_ratio_chi_m}
	\chi_\text{v,mm}
	=\left(1+\frac{\chi_\text{mm}^{xx}}{d}\right)\frac{H_{x,\text{av}}^\text{m}}{H_{x,\text{av}}^\text{s}}-1.
	\end{equation}
\end{subequations} \textbf{}
These relations~\eqref{eq:av_ratio_chi} represent the general and exact equivalence relations between the surface susceptibilities ($\chi$) of the metasurface in Fig.~\ref{Fig:diluted_ms_approx}a and the voluminal susceptibility ($\chi_\text{v}$) of the slab in Fig.~\ref{Fig:diluted_ms_approx}b, which naturally involve the thickness of the slab, but also the average fields across the metasurface and the average fields across the slab. Unfortunately, these relations are \emph{implicit}: the volume susceptibilities depend on the average slab fields, which obviously depend themselves on the volume susceptibilities. So, they are not practical per se. However, they will serve as the basis for the Average Field Approximation (AFA) in Sec.~\ref{sec:AFA}.

\section{Explicit Exact Relations}\label{sec:exp_ex_rel}
\subsection{Derivations}\label{sec:derivations}
We shall now follow another approach to find an \emph{explicit} alternative to the implicit relations~\eqref{eq:av_ratio_chi}. This approach will consist two steps: 1)~equating the scattering coefficients of the metasurface and of the thin slab, and 2)~solving the resulting equations fo the volume susceptibilities in terms of the metasurface susceptibilities and other slab parameters.

The relations between the bianisotropic metasurface susceptibilities and scattering coefficients are available in closed-form for most metasurfaces~\cite{Achouri_general_ms_synthes_2015,Achouri_NP_06_2018,Achouri_EMTA_2020}. Under the assumptions of this paper (Sec.~\ref{sec:sta_prob}), these are found by substituting~\eqref{eq:EH_difav} into~\eqref{eq:2D_GSTC}  as
\begin{subequations}\label{eq:suscep_ms_normal}
	\begin{equation}
	\chi_\text{ee}^{yy}=\frac{2}{jk_0} \frac{1-\Gamma -T} {1+\Gamma +T},
	\quad
	\chi_\text{mm}^{xx}=\frac{2}{jk_0} \frac{1+\Gamma -T} {1-\Gamma +T},
	\end{equation}
\end{subequations}
or, inversely, 
\begin{subequations}\label{eq:TG_suscep}
	\begin{equation}\label{eq:TG}
	T=\frac{1}{2}\left(\frac{1-p}{1+p}+\frac{1-q}{1+q}\right),
	\;
	\Gamma=\frac{1}{2}\left(\frac{1-p}{1+p}-\frac{1-q}{1+q}\right),
	\end{equation}
with
\begin{equation}\label{eq:pq}
p={jk_0}\chi_\text{ee}^{yy}/2
\quad\text{and}\quad
q = {jk_0} \chi_\text{mm}^{xx}/2,
\end{equation}
\end{subequations}
where $T$ and $\Gamma$ are the transmission coefficient and the reflection coefficients, respectively, and where $k_0=\omega/c$.

On the other hand, the exact relations between the susceptibilities and scattering coefficients of a slab read~\cite{jinau2000electromagnetic}
\begin{subequations}\label{eq:slab_trans_ref}
	\begin{equation}\label{eq:slab_TG}
	T=\frac{\left( 1-R^2\right)\text{e}^{-j(k-k_0)d}}{1-R^2\text{e}^{-2jkd}},
	\;
	\Gamma=\frac{R\text{e}^{jk_0d}\left(\text{e}^{-j2kd}-1\right)}{1-R^2e^{-2jkd}},
	\end{equation}
	where
	\begin{equation}\label{eq:RR}
	R=\frac{1-\eta_\text{r}}{1+\eta_\text{r}}
	\end{equation}
	with
	\begin{equation}\label{eq:etar}
	\eta_\text{r}=\sqrt{\frac{\mu_\text{r}}{\epsilon_\text{r}}}
	\end{equation}	
	and where
	\begin{equation}\label{eq:kr}
	k=k_0\sqrt{\epsilon_\text{r}\mu_\text{r}}
	\end{equation}
	with
	\begin{equation}\label{eq:ermr}
	\epsilon_\text{r}=1+\chi_\text{v,ee}
	\quad\text{and}\quad
	\mu_\text{r}=1+\chi_\text{v,mm}.
	\end{equation}
\end{subequations}
The reverse relations are found by solving~\eqref{eq:etar} and~\eqref{eq:kr} for ($\epsilon_\text{r}$, $\mu_\text{r}$) in terms of ($\eta_\text{r}$, $k$), replacing $\eta_\text{r}$ in the two resulting equations by $(1-R)/(1+R)$ following~\eqref{eq:RR}, and finally using~\eqref{eq:ermr}. This yields
\begin{subequations}\label{eq:inv_R_k}
		\begin{equation}\label{eq:ex_rel}
	\chi_{\text{v,ee}}=\frac{k}{k_0}\left(\frac{1+R}{1-R}\right)-1,
\;
	\chi_{\text{v,mm}}=\frac{k}{k_0}\left(\frac{1-R}{1+R}\right)-1,
	\end{equation}   
where $R$ and $e^{-jkd}$ can be found be inverting~\eqref{eq:slab_TG} as
	\begin{equation}\label{eq:R}
	R=\frac{-1+\tau^2-\gamma^2+\sqrt{(1-\tau^2+\gamma^2)^2-4\gamma^2}}{2\gamma},
	\end{equation}
	\begin{equation}\label{eq:exp_kd}
	e^{-jkd}=\frac{1+\tau^2-\gamma^2-\sqrt{(1-\tau^2+\gamma^2)^2-4\gamma^2}}{2\tau},
	\end{equation}
	with
	\begin{equation}\label{eq:taugam}
	\tau=Te^{-jk_0d}
	\quad\text{and\quad}
	\gamma=\Gamma e^{-jk_0d}.
	\end{equation}
\end{subequations}

The sought after exact explicit relations are finally obtained by equating the metasurface and slab scattering coefficients, according to the following sequence:
\begin{gather}\label{eq:procedure}
\chi_\text{ee}^{yy},\chi_\text{mm}^{xx}
\stackrel{\eqref{eq:TG_suscep}}{\longrightarrow}
T^\text{m},\Gamma^\text{m}=T^\text{s},\Gamma^\text{s}
\stackrel{\eqref{eq:inv_R_k}}{\longrightarrow}
\chi_{\text{v,ee}},\chi_{\text{v,mm}}.
\end{gather}

\subsection{Results}\label{sec:results}
\begin{figure}[!ht]
	\centering
	\begin{subfigure}{0.48\columnwidth}
		\centering
		\includegraphics[width=1\columnwidth]{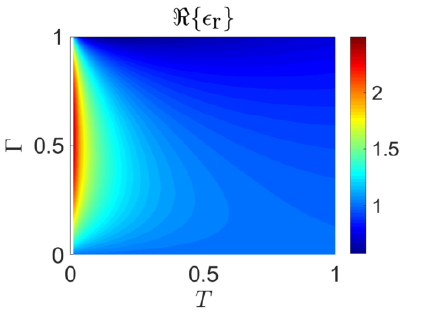}
		\caption{}\label{FIG:real_epsr}
	\end{subfigure}
	\begin{subfigure}{0.48\columnwidth}
		\centering
		\includegraphics[width=1\columnwidth]{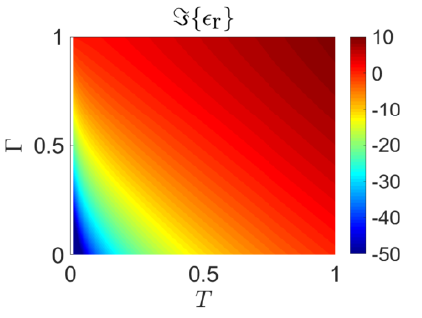}
		\caption{}\label{Fig:imag_epsr}
	\end{subfigure}
	
	\begin{subfigure}{0.48\columnwidth}
		\centering
		\includegraphics[width=1\columnwidth]{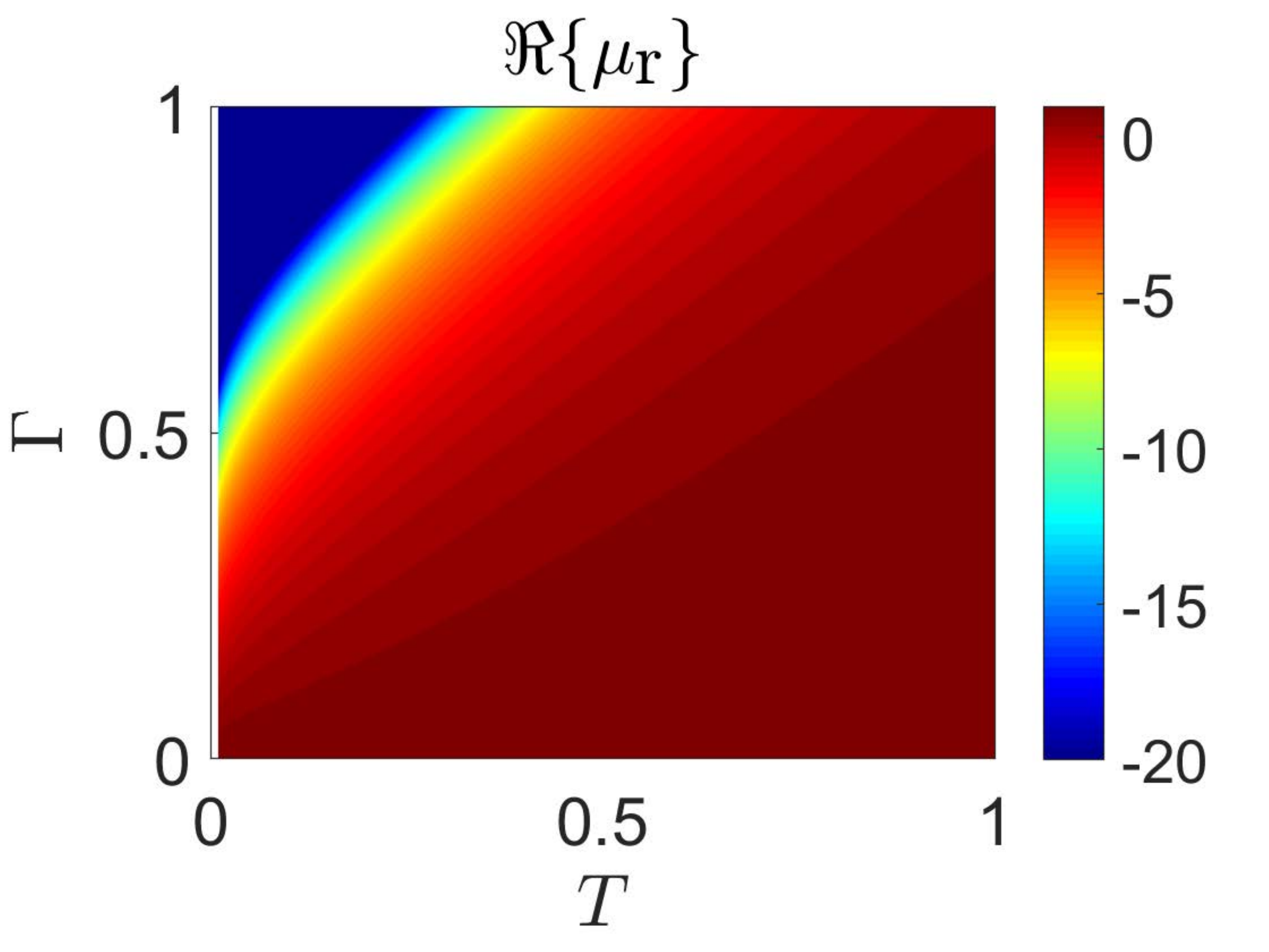}
		\caption{}\label{Fig:real_mur}
	\end{subfigure}
	\begin{subfigure}{0.48\columnwidth}
		\centering
		\includegraphics[width=1\columnwidth]{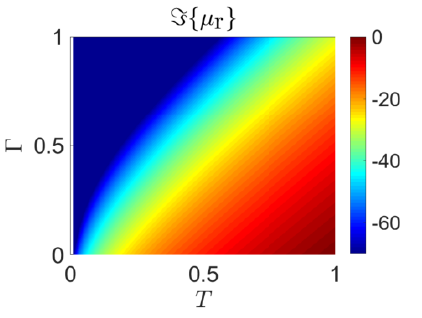}
		\caption{}\label{Fig:imag_mur}
	\end{subfigure}
	\caption{Exact slab parameters $\epsilon_\text{r}$ and $\mu_\text{r}$ versus $T$ and $\Gamma$ using~\eqref{eq:inv_R_k} and~\eqref{eq:ermr} for $d=\lambda_0/100$. (a)~$\Re(\epsilon_\text{r})$. (b)~$\Im(\epsilon_\text{r})$. (c)~$\Re(\mu_\text{r})$. (d)~$\Im(\mu_\text{r})$.} \label{Fig:epsr_mur}
\end{figure}

\begin{figure}[!ht]
	\centering
	\begin{subfigure}{0.48\columnwidth}
		\centering
		\includegraphics[width=1\columnwidth]{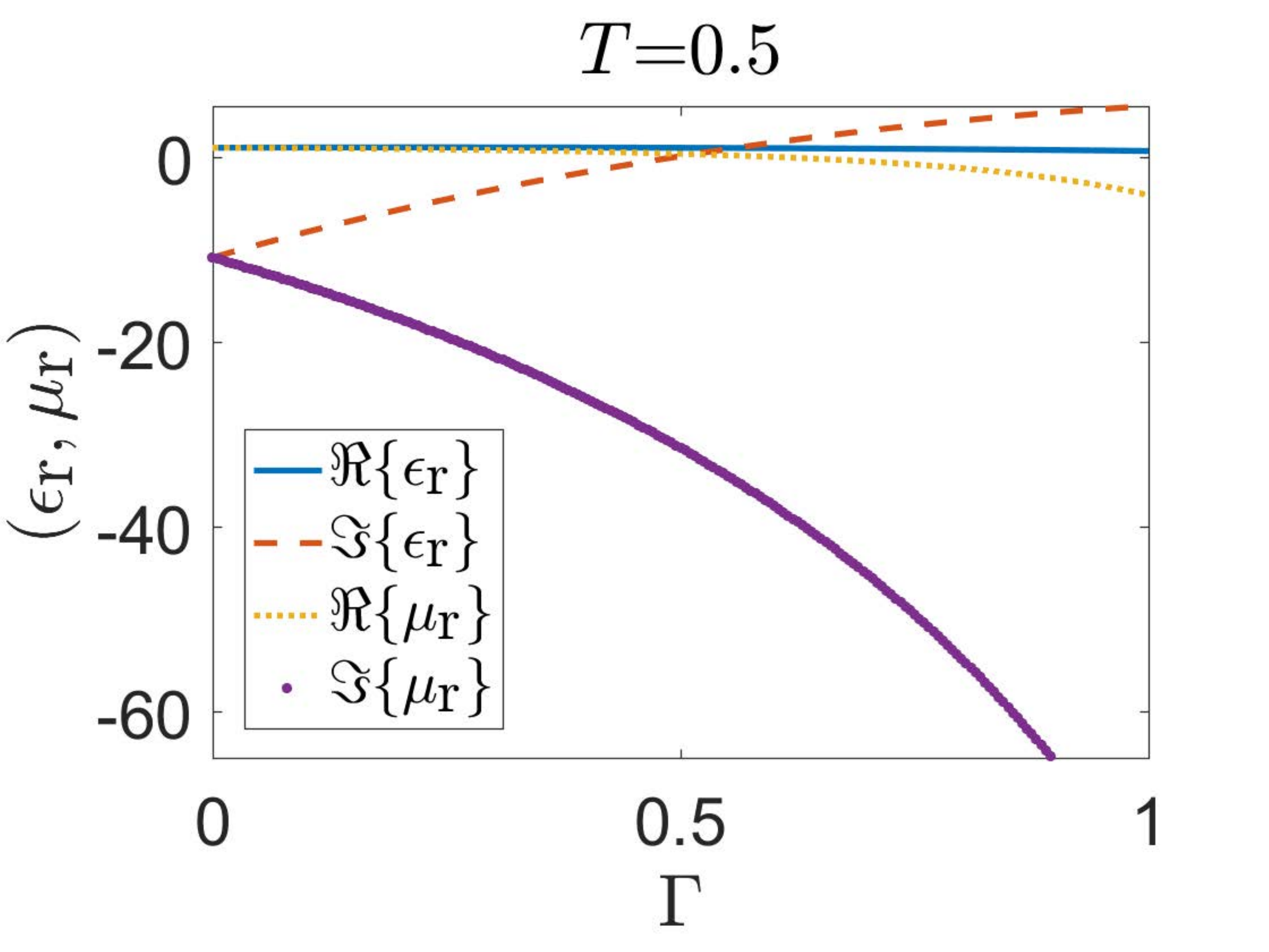}
		\caption{}\label{FIG:par_Tfixed}
	\end{subfigure}
	\begin{subfigure}{0.48\columnwidth}
		\centering
		\includegraphics[width=1\columnwidth]{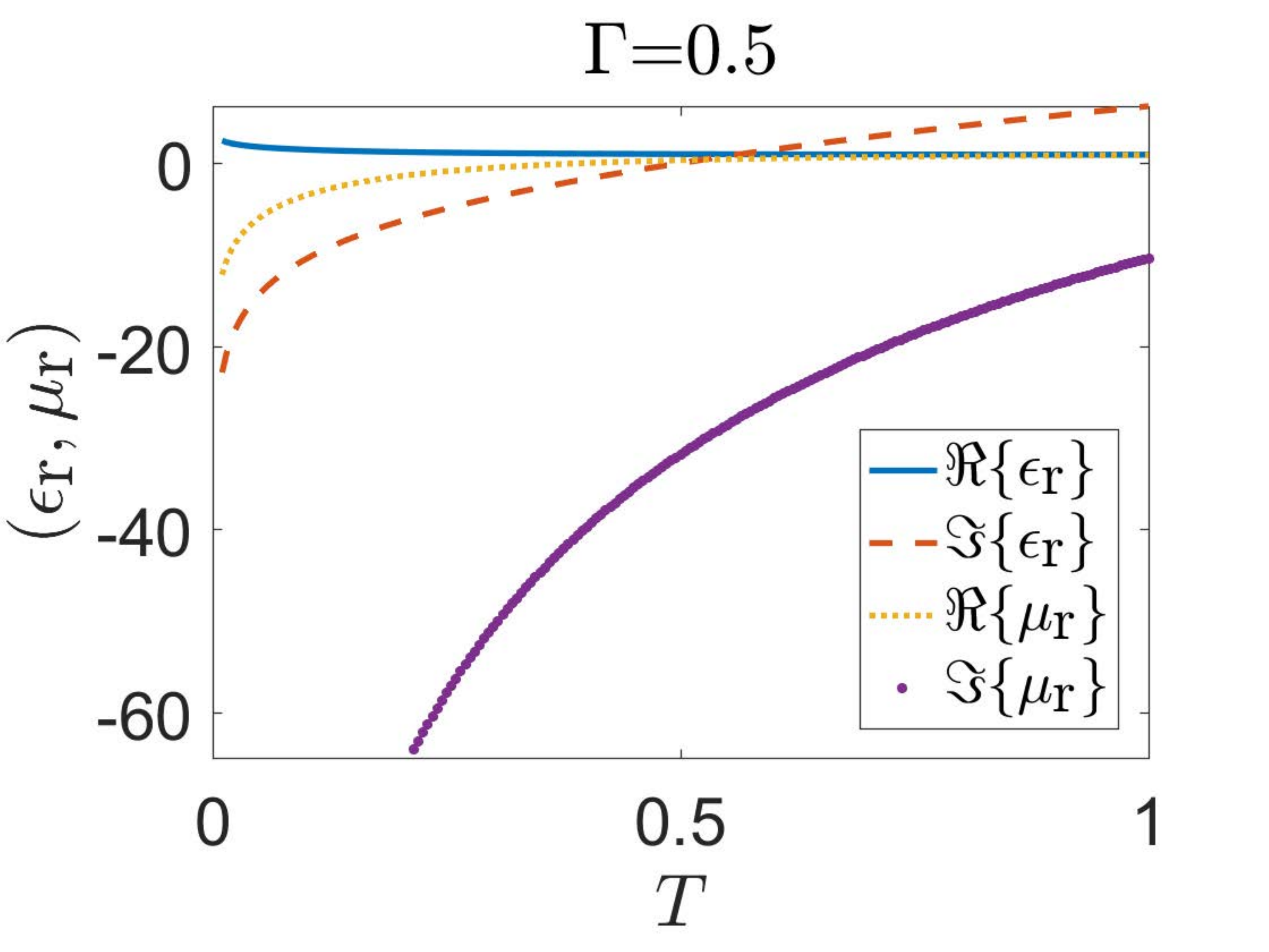}
		\caption{}\label{FIG:par_Gfixed}
	\end{subfigure}
	\caption{Cuts of Fig.~\ref{Fig:epsr_mur} at (a)~$T=0.5$ and (b)~$\Gamma=0.5$.} \label{Fig:epsr_mur_T_G}
\end{figure}

Figure~\ref{Fig:epsr_mur} plots the real and imaginary parts of the exact slab parameters $\epsilon_\text{r}$ and $\mu_\text{r}$ versus $T$ and $\Gamma$ using~\eqref{eq:inv_R_k} and~\eqref{eq:ermr} for a fixed value of $d$, while Fig.~\ref{Fig:epsr_mur_T_G} shows two  cuts in Fig.~\ref{Fig:epsr_mur}. These figures show that scanning the $(T,\Gamma)$ plane from $0$ to $1$ in each direction requires a great diversity of nontrivial positive and negative real and negative parameters.

\begin{figure}[!ht]
	\centering
	\begin{subfigure}{0.48\columnwidth}
		\centering
		\includegraphics[width=1\columnwidth]{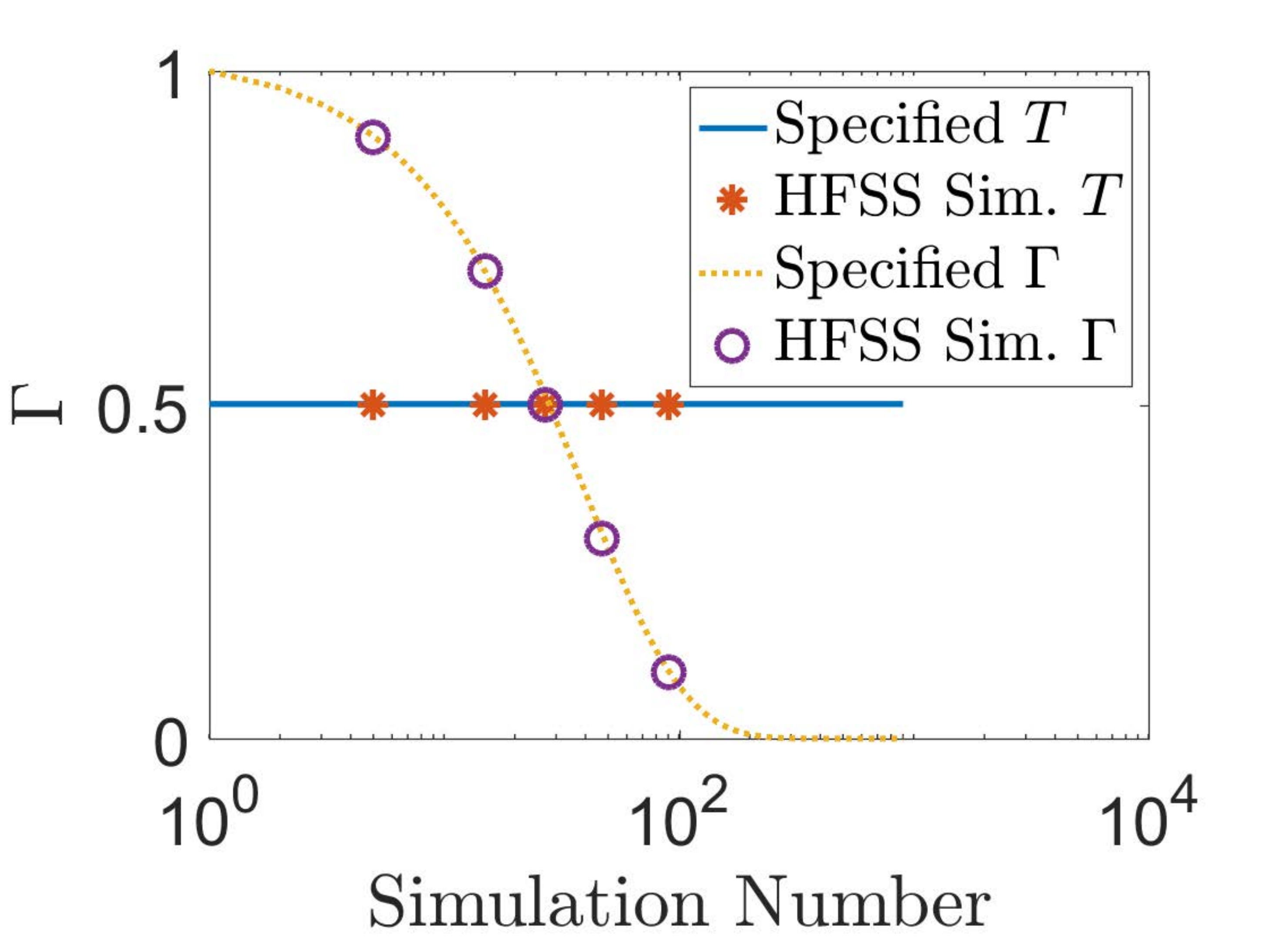}
		\caption{}\label{Fig:Tfixed}
	\end{subfigure}
	\begin{subfigure}{0.48\columnwidth}
		\centering
		\includegraphics[width=1\columnwidth]{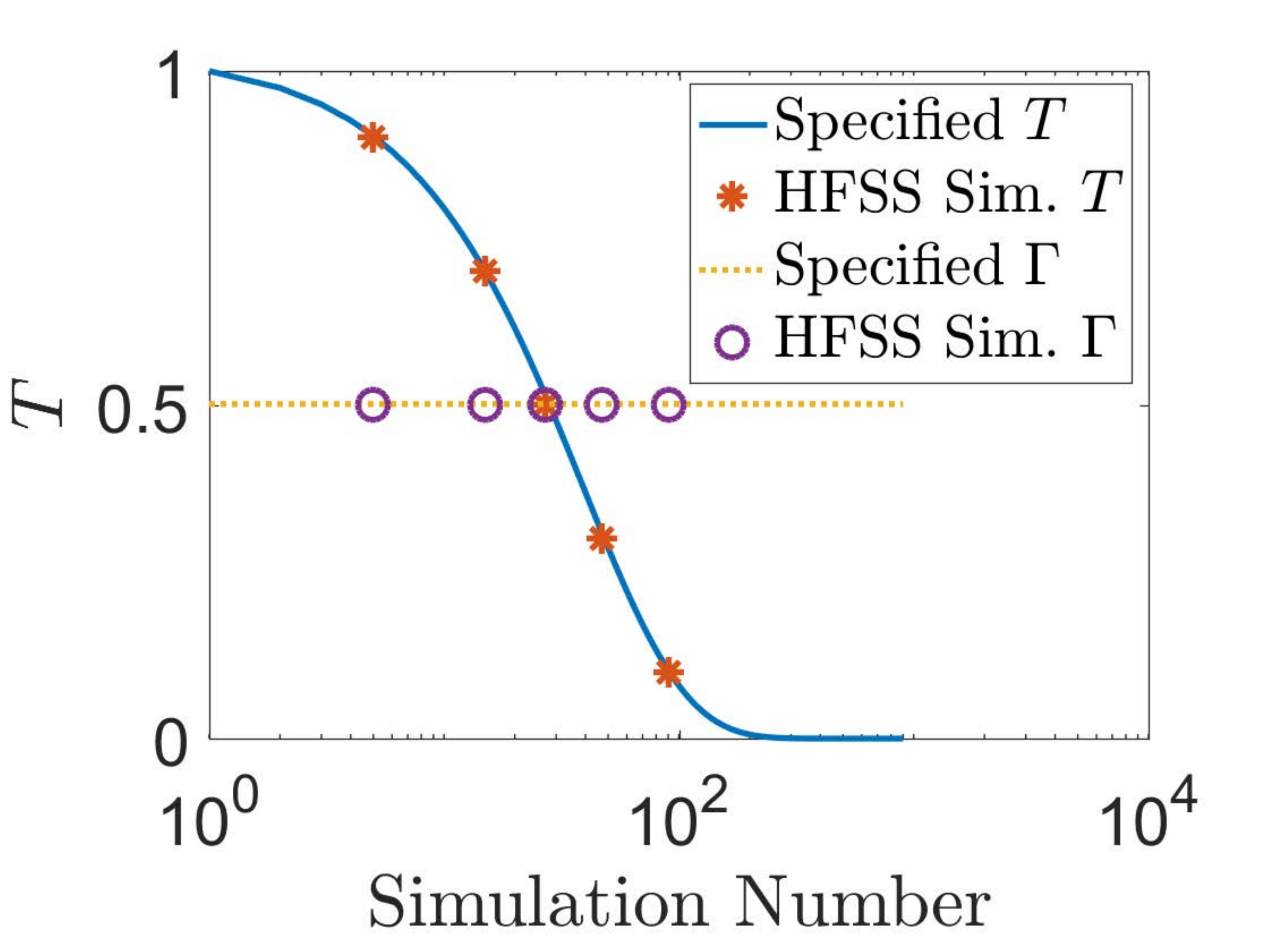}
		\caption{}\label{Fig:Gfixed}
	\end{subfigure}
	\caption{Full-wave (HFSS) validation of the procedure~\eqref{eq:procedure}, still for $d=\lambda_0/100$, substituting the closed-form metasurface scattering parameters~\eqref{eq:TG_suscep} into~\eqref{eq:inv_R_k} for metasurface susceptibilities values spanning (a)~ $\Gamma\in[0,1]$ at $T=0.5$ and (b)~$T\in[0,1]$ at $\Gamma=0.5$.} \label{Fig:agreement}
\end{figure}

We shall now validate the procedure given in~\eqref{eq:procedure}. For this purpose, we first calculate the thin-slab voluminal susceptibility (or permittivity and permeability) values for given specified scattering parameters by injecting the values of these parameters, $T$ and $\Gamma$, into~\eqref{eq:inv_R_k} [specifically, into~\eqref{eq:taugam}]. Then we full-wave simulate the slab having these parameter values using a commercial software. Finally, we check whether the scattering coefficients $S_{21}$ and $S_{11}$ obtained by the simulator match the specified $T$ and $\Gamma$ values. Figure~\ref{Fig:agreement} plot the result the same fixed parameter $T$ or $\Gamma$ as in Fig.~\eqref{Fig:epsr_mur_T_G}. The perfect agreement between the simulation and the specification confirms the correctness of the exact analytical formula~\eqref{eq:inv_R_k} and the validity of the overall proposed procedure.

As an application example, let us now apply our procedure to a matched ($\Gamma=0$) metasurface attenuator of power attenuation $A=1-T^2$. The corresponding surface susceptibilities are found by substituting $\Gamma=0$ in~\eqref{eq:suscep_ms_normal} as
\begin{equation}\label{eq:suscep_aborb_ms}
\chi_\text{ee}^{yy}=\chi_\text{mm}^{xx}=\frac{2j}{k_0}\frac{T-1}{T+1}, 
\end{equation}
which are noted to be purely imaginary, consistently with the specification of absorption~\cite{Achouri_EMTA_2020}. The corresponding voluminal susceptibilities are found from~\eqref{eq:inv_R_k} with $R=0$ as $\chi_{\text{v,ee}}=\chi_{\text{v,mm}}=k/k_0-1=(k-k_0)/k_0$ or, in terms of $T$, using~\eqref{eq:slab_TG} (still with $R=0$), as
\begin{equation}\label{eq:suscep_aborb_s}
\chi_{\text{v,ee}}=\chi_{\text{v,mm}}=j\frac{\ln T}{k_0 d}.
\end{equation}

\begin{figure}[!ht]
	\centering
	\includegraphics[width=1\columnwidth]{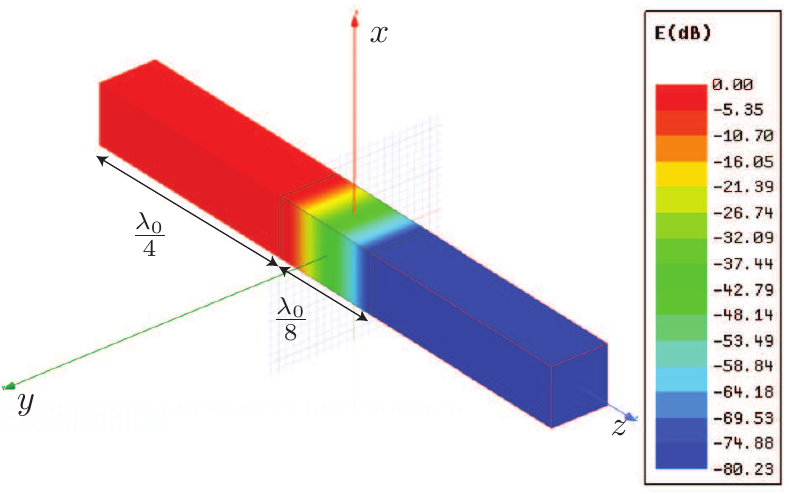}{
    }
	\caption{Full-wave simulation (HFSS) of a thin slab illuminated by a plane wave, with the slab parameters computed by~\eqref{eq:suscep_aborb_s} for modeling a quasi-perfectly absorbing metasurface with $T=10^{-4}$ ($S_{21}=-80$~dB) and slab width $d={\lambda_0}/{8}$, i.e., $\chi_\text{v,ee}=\chi_\text{v,mm}=-j11.73$ at $1$~GHz, corresponding to the metasurface susceptibilities $\chi_\text{ee}^{yy}=\chi_\text{mm}^{xx}\approx-j0.096$ according to~\eqref{eq:suscep_aborb_ms}.}\label{Fig:Slab_ms_exact_result}
\end{figure}

Figure~\ref{Fig:Slab_ms_exact_result} shows the full-wave simulated field for a quasi-perfectly absorbing thin slab with the voluminal susceptibility given by~\eqref{eq:suscep_aborb_s}. The result verifies that the slab with the explicit slab parameters~\eqref{eq:inv_R_k} behaves essentially as the metasurface it should model, the attenuation level being extremely close ($80.23$~dB) to the prescribed level ($80$~dB).

Interestingly, for $T\rightarrow{1}$, Eq.~\eqref{eq:suscep_aborb_ms} reduces to $\chi\approx{j}(T-1)/(k_0)$ while Eq.~\eqref{eq:suscep_aborb_s} reduces to $\chi_\text{v}\approx j(T-1)(k_0 d)$, so that in this particular case the two susceptibilities are simply related as $\chi_\text{v}=\chi/d$. We will further discuss this relation in Sec.~\ref{sec:AFA}.

\section{Average Field Approximation (AFA)}\label{sec:AFA}

\subsection{Motivation}

The relations~\eqref{eq:inv_R_k} between ($\chi_\text{ee}^{yy}$,$\chi_\text{mm}^{xx}$) and ($\chi_{\text{v,ee}}$,$\chi_{\text{v,mm}}$) are explicit in the sense that the procedure~\eqref{eq:procedure}, which is perfectly convenient for a computer routine, seamlessly relates the former to the latter. However, that explicit relation is not explicitly \emph{written} in Sec.~\ref{sec:derivations} as $(\chi_\text{ee}^{yy},\chi_\text{mm}^{xx})=f[(\chi_{\text{v,ee}},\chi_{\text{v,mm}})]$ because the function $f(\cdot)$, given the mediation of the scattering parameters, would be very cumbersome\footnote{It would imply the following chain of analytical substitutions: 
$\chi_\text{ee}^{yy},\chi_\text{mm}^{xx}\stackrel{\eqref{eq:pq}}{\longrightarrow}
p,q\stackrel{\eqref{eq:TG}}{\longrightarrow}
T,G\stackrel{\eqref{eq:taugam}}{\longrightarrow}
\tau,\gamma\stackrel{\eqref{eq:R},\eqref{eq:exp_kd}}{\longrightarrow}
R,d\stackrel{\eqref{eq:ex_rel}}{\longrightarrow}
\chi_{\text{v,ee}},\chi_{\text{v,mm}}$.}. It would therefore be profitable to establish, if possible, an \emph{approximate} formula that would be both more handy and more insightful. This section derives such a formula, illustrates its restrictions and derives its range of validity.

\subsection{AFA Formula}

The simplest approximation approach consists in assuming that the field across the slab in Fig.~\ref{Fig:diluted_ms_approx}b is \emph{uniform}, which we call here the Average Field Approximation (AFA) approach. The AFA seems a priori reasonable given the subwavelength nature of the slab, although we know that it is strictly incorrect: the electromagnetic field \emph{does} vary across the slab, except in the trivial case of a metasurface with vanishingly small susceptibilities or complete transmission (limit of no metasurface). Therefore, the AFA approximation is expected to be satisfactory only in the range of parameters where the actual equivalent field variation across the slab is negligibly small.

Setting the field average across the slab to a constant equal to the average of the fields at the two edges of the slab, which is identical to the field average across the metasurface in~\eqref{eq:E_difava}, i.e.,
\begin{equation}\label{eq:AFA_approx}
E_{y,\text{av}}^\text{s}\approx{}E_{y,\text{av}}^\text{m},
\end{equation}
reduces~\eqref{eq:av_ratio_chi_e} to
\begin{equation}\label{eq:solution}
\chi_\text{v,ee}\approx\frac{\chi_\text{ee}^{yy}}{d}.
\end{equation}
This extremely simple formula is the AFA approximation formula. It may be interpreted as uniformly distributing the effect of the metasurface across the slab. According to the last paragraph of Sec.~\eqref{sec:results}, this general approximation tends to be exact in the limit situation where the metasurface vanishes ($\Gamma,T$), as should be the case. 

The AFA formula [Eq.~\eqref{eq:solution}] has been successfully applied to relatively simple metasurfaces in the literature~\cite{Achouri_improv_ms_discon_cond_2015,Liu_homogenized_MS_2018}, but it should unavoidably fail when the fields strongly vary across the slab, or the metasurfaces to model involve strong field transformations.

\subsection{Failure Example}\label{sec:intuition}
To illustrate the restriction of the AFA approximation, let us come back to the matched metasurface attenuator discussed in Sec.~\ref{sec:results}. Inserting.~\eqref{eq:suscep_aborb_ms} into~\eqref{eq:solution} gives the AFA result
\begin{equation}\label{eq:AFA_approximation}
\chi_\text{v,ee}=\chi_\text{v,mm}\approx\frac{2j}{k_0d}\frac{T-1}{T+1}.
\end{equation}

\begin{figure}[!h]
	\centering
	\includegraphics[width=1\columnwidth]{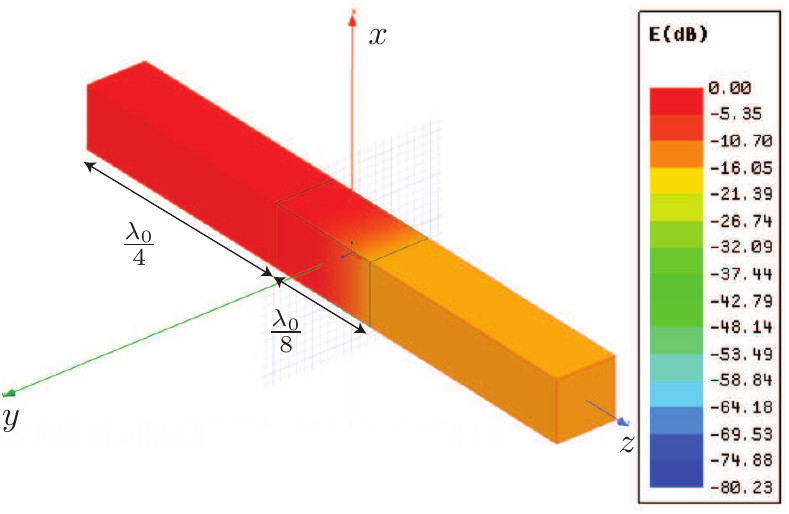}{
    }
	\caption{Failure of the AFA formula~\eqref{eq:solution} to model a perfectly absorbing  metasurface (still with $d={\lambda_0}/{8}$): $T\approx{0.135}$ ($S_{21}\approx-17.4$~dB).}\label{FIG:ABSabsorber_COMSOL}
\end{figure}

Figure~\ref{FIG:ABSabsorber_COMSOL} shows the full-wave simulated field for a perfectly absorbing thin slab with the AFA voluminal susceptibility given by~\eqref{eq:AFA_approximation} for the  case of perfect absorption ($T=0$, in addition to $\Gamma=0$). In contrast to the exact formula (Fig.~\ref{Fig:Slab_ms_exact_result}), the AFA formula clearly fails to model the absorbing metasurface, since the structure passes almost $15\%$ of the power whereas it should transmit nothing! This example illustrates the fact that the AFA formula unavoidably fails in the case of extreme metasurface field transformations.

\begin{figure}[!h]
	\centering
	\includegraphics[width=1\columnwidth]{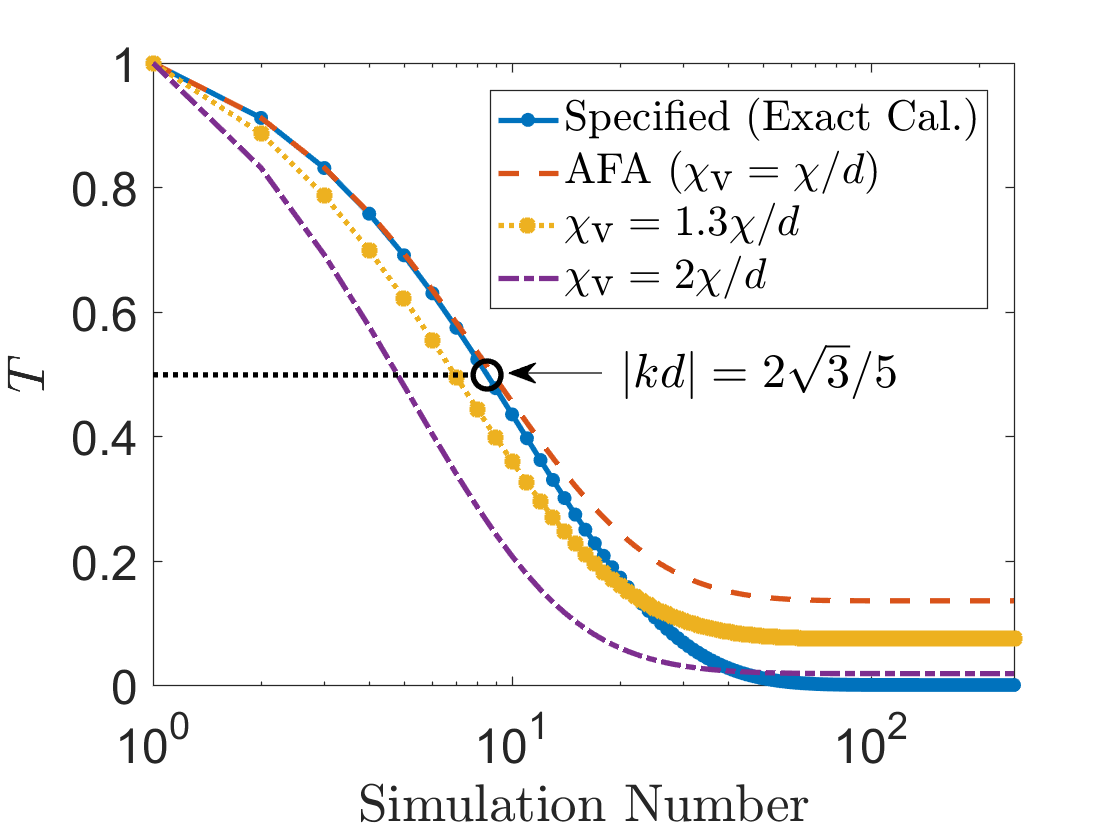}
	\caption{Full-wave simulation results for the transmission across the slab using the AFA susceptibility $\chi_\text{v}=\chi/d$ [Eq.~\eqref{eq:solution}] with $d=\lambda/100$ to model a reflectionless absorbing metasurface with transmission coefficient $T$~\cite{Achouri_improv_ms_discon_cond_2015}. The curves obtained for $\chi_\text{v}=1.3\chi/d$ and $\chi_\text{v}=2\chi/d$ are also shown for the discussion in Sec.~\ref{sec:intuition} and Sec.~\ref{sec:Validity_Range}.}\label{Fig:reflectionless_mt_parametric}
\end{figure}

Figure~\ref{Fig:reflectionless_mt_parametric} compares the full-wave simulated transmission coefficients of a slab with the approximated voluminal susceptibility~\eqref{eq:AFA_approximation} and a deeply subwavelength thickness $d=\lambda_0/100$ with the specified (or exactly calculated) transmission coefficients. While closely following the metasurface specification at relatively high transmission (or low absorption) levels, the AFA deviates more and more from the specification as $T$ decreases below $0.5$ (or $A$ increases above $75\%$). We shall show in Sec.~\ref{sec:Validity_Range} that this point $T=0.5$ corresponds to $|kd|=2\sqrt{3}/5$. 

\begin{figure}[h]
	\centering
	\begin{subfigure}{0.49\columnwidth}
		\centering
		\includegraphics[width=1\columnwidth]{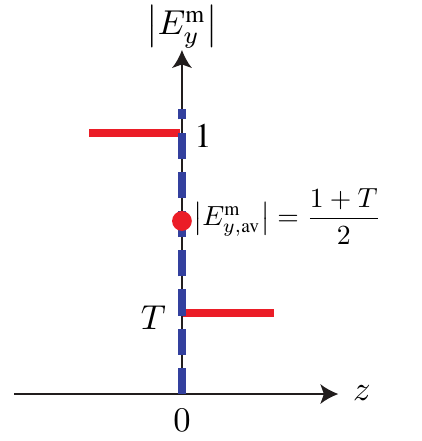}{
		}\caption{}\label{FIG:MS_slab_compare_1}
	\end{subfigure}
	\begin{subfigure}{0.49\columnwidth}
		\centering
		\includegraphics[width=1\columnwidth]{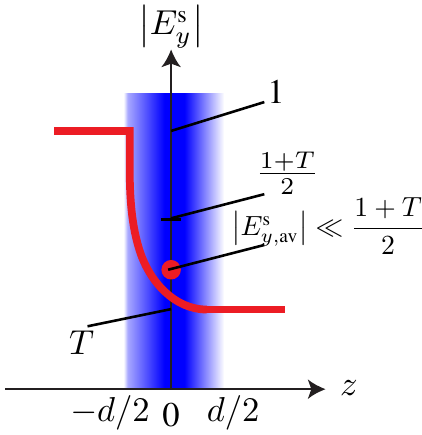}{
		} \caption{}\label{Fig:MS_slab_compare_2}
	\end{subfigure}
	\caption{Difference between the metasurface and slab field distributions and resulting average fields for the absorber in Fig.~\ref{Fig:reflectionless_mt_parametric}. (a)~Metasurface (Fig.~\ref{Fig:diluted_ms_approx}a). (b)~Slab (Fig.~\ref{Fig:diluted_ms_approx}b).} \label{Fig:MS_slab_compare}
\end{figure}

The reason for this discrepancy is the fact that, as the absorption progressively increases, the actual field average across the slab, $E_{y,\text{av}}^{s}$, given in Appendix~A by Eq.~\eqref{eq:Eyavs_evaluated}, increasingly deviates from the average of the fields at either side of the metasurface, $E_{y,\text{av}}^\text{m}$, given by~\eqref{eq:E_difava}, as illustrated in Fig.~\ref{Fig:MS_slab_compare}. Then the approximation~\eqref{eq:AFA_approx} is not valid anymore, and therefore the AFA formula~\eqref{eq:solution} is inapplicable. Taking into account this difference in the field averages would naturally lead to a better approximation for $\chi_\text{v,ee}$, corresponding to the more accurate formula.

Increasing the specified absorption implies increasing the metasurface susceptibility, $\chi_\text{ee}^{yy}$, since this corresponds to increasing the effect of the metasurface on the incident wave. At the same time, increasing the absorption implies decreasing $E_{y,\text{av}}^\text{s}$, according to the rationale of Fig.~\ref{Fig:MS_slab_compare}. Therefore, increasing the absorption requires, according to~\eqref{eq:av_ratio_chi}, increasing the susceptibility ($\chi_\text{v,ee}$) by the factor $\chi_\text{ee}^{yy}/E_{y,\text{av}}^\text{s}$, which the simple formula~\eqref{eq:solution} fails to accomplish. This results in a too small susceptibility ($\chi_\text{v}$) and hence in the too high transmission level observed in Figs.~\ref{FIG:ABSabsorber_COMSOL} and~\ref{Fig:reflectionless_mt_parametric}. 

However, this underestimation of the susceptibility~\eqref{eq:solution} is not all of its limitation. As shown in Fig.~\ref{Fig:reflectionless_mt_parametric}, the level of transmission of the AFA model saturates at some level (i.e., about $T=0.135$). Increasing $\chi_\text{v}$ in the AFA approach would better model highly absorptive metasurfaces. For example as shown in Fig.~\ref{Fig:reflectionless_mt_parametric}, magnifying the susceptibilities by $30\%$ and $100\%$ ends up to smaller transmissions (i.e., respectively $T=0.07$ and $T=0.01$). However, this modification loses its validity to model metasurfaces with higher transmission coefficients values. 

\subsection{Validity Range of the AFA}\label{sec:Validity_Range}

The observations of the previous subsections prompts for the determination of a range of validity for the AFA model [Eq.~\eqref{eq:solution}]. For this purpose, we have to determine the condition under which $E_{y,\text{av}}^\text{s}\approx{}E_{y,\text{av}}^\text{m}$ and $H_{x,\text{av}}^\text{s}\approx{}H_{x,\text{av}}^\text{m}$ in~\eqref{eq:av_ratio_chi}, i.e., under which condition the average expressions in~\eqref{eq:Eyavs} and~\eqref{eq:E_difava} [\eqref{eq:Hxavs}~and~\eqref{eq:H_difava}] are approximately equal.

\subsubsection{Case of the Matched Metasurface Attenuator}

Let us first examine the matched metasurface attenuator, for which the exact voluminal susceptibilities are given by~\eqref{eq:suscep_aborb_s}. In this case, $R=0$, and therefore $\eta_\text{r}=1$ according to~\eqref{eq:slab_trans_ref}. The expression for the exact average electric field in the slab [Eq.~\eqref{eq:integ_field_slab} or Eq.~\eqref{eq:step2_simplification} in Appendix~A] reduces to
\begin{equation}\label{eq:integ_field_slab2}
E_{y,\text{av}}^\text{s}
=\frac{2A}{kd}\sin\left(\frac{kd}{2}\right),
\end{equation}
where to $A=\text{e}^{-j\left(k-k_0\right)d/2}$. However, since the slab is subwavelength ($\text{e}^{jk_0d/2}\approx{1}$), we have $A\approx\text{e}^{-jkd/2}$. On the other hand, the average field across the metasurface is found from~\eqref{eq:EH_difav} with $E_{y}^\text{i}=1$, $E_{y}^\text{r}=0$ and $E_{y}^\text{t}=T$, as
\begin{equation}\label{eq:EY_av_1}
E_{y,\text{av}}^\text{m}=\frac{1+T}{2}.
\end{equation}
Since we have from $R=0$ that $T=\text{e}^{-j(k-k_0)d}$, and hence $T\approx\text{e}^{-jkd}$, Eq.~\eqref{eq:EY_av_1} reduces to
\begin{equation}\label{eq:EY_av}
E_{y,\text{av}}^\text{m}\approx A\cos\left(\frac{kd}{2}\right).
\end{equation}

Equating the slab average electric field~\eqref{eq:integ_field_slab2} with the approximate metasurface average electric field~\eqref{eq:EY_av} yields
\begin{equation}\label{eq:equating}
\tan\left(\frac{kd}{2}\right) \approx \frac{kd}{2},
\end{equation}
and similarly equating the average magnetic fields may be easily verified to yield the same relation. Given the subwavelength thickness of the slab, i.e., $|kd|\ll{1}$, we may approximate the tangent function by its Taylor series approximation $\tan xa\\prox x+\frac{1}{3}x^3+\ldots$, which simplifies~\eqref{eq:equating} to
\begin{subequations}
	\begin{equation}
	\frac{kd}{2}+\frac{1}{3}\left(\frac{kd}{2}\right)^3\approx\frac{kd}{2},
	\end{equation}
	or
	\begin{equation}
	1+\frac{1}{3}\left(\frac{kd}{2}\right)^2\approx{1},
	\end{equation}
	which implies
	\begin{equation}
	\frac{1}{3}\left|\frac{kd}{2}\right|^2\ll{1},
	\end{equation}
	i.e.,
	\begin{equation}\label{eq:approximated_kd}
	|kd|\ll 2\sqrt{3}\approx 3.4841.
	\end{equation}
\end{subequations}
This relation represents thus the range of validity of the AFA for the case of the matched metasurface attenuator. Taking a marging factor of $5$, we therefore expect the AFA to be a good approximation for $|kd|<2\sqrt{3}/5 \approx 0.7$. We will next show that $|kd| \approx 0.7$ corresponds to $T \approx 0.5$, which is consistent with the result plotted in Fig.~\ref{Fig:reflectionless_mt_parametric}.  

\subsubsection{General Case}

Let us now consider the general case where $T$ and $\Gamma$ are both nonzero, i.e., let us determine under which general condition $E_{y,\text{av}}^\text{s}\approx{}E_{y,\text{av}}^\text{m}$ and $H_{x,\text{av}}^\text{s}\approx{}H_{x,\text{av}}^\text{m}$ in this case. For this purpose, we compare the exact slab average fields by~\eqref{eq:step2_simplification} from the exact voluminal susceptibilities given by~\eqref{eq:inv_R_k}, for the safely subwavelength thickness of $d=\lambda_0/100$, and the exact metasurface average fields given here by~\eqref{eq:E_difava} and~\eqref{eq:H_difava} as $E_{y,\text{av}}^\text{m}=(1+\Gamma+T)/2$ and $H_{x,\text{av}}^\text{m}=(-1+\Gamma-T)/(2\eta_0)$, versus $(T,\Gamma)$.

\begin{figure}[!h]
	\centering
	\begin{subfigure}{0.48\columnwidth}
		\centering
		\includegraphics[width=1\columnwidth]{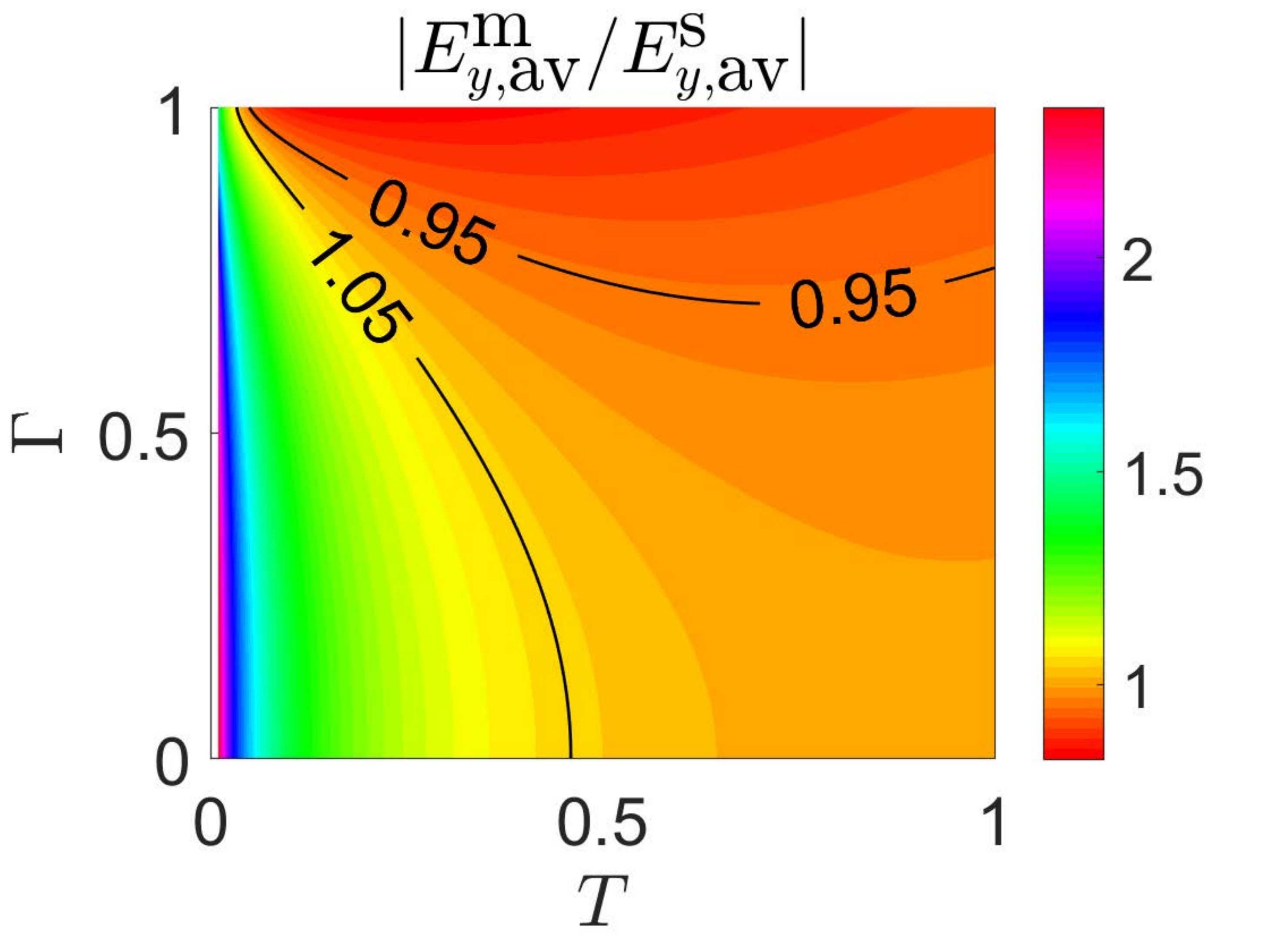}
		\caption{}\label{Fig:Ratio_E}
	\end{subfigure}
	\begin{subfigure}{0.48\columnwidth}
		\centering
		\includegraphics[width=1\columnwidth]{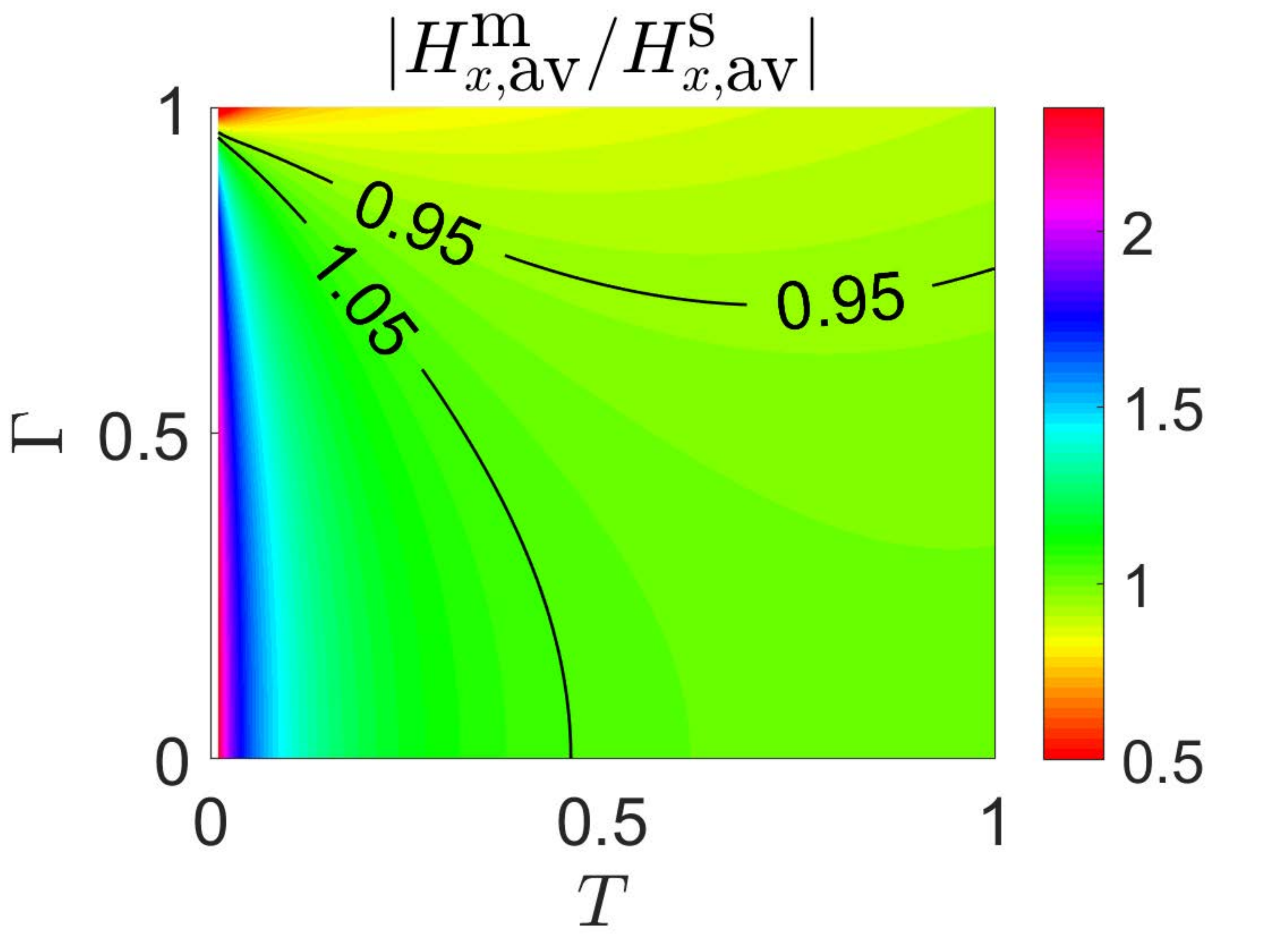}
		\caption{}\label{Fig:Ratio_H}
	\end{subfigure}
	\begin{subfigure}{0.48\columnwidth}
		\centering
		\includegraphics[width=1\columnwidth]{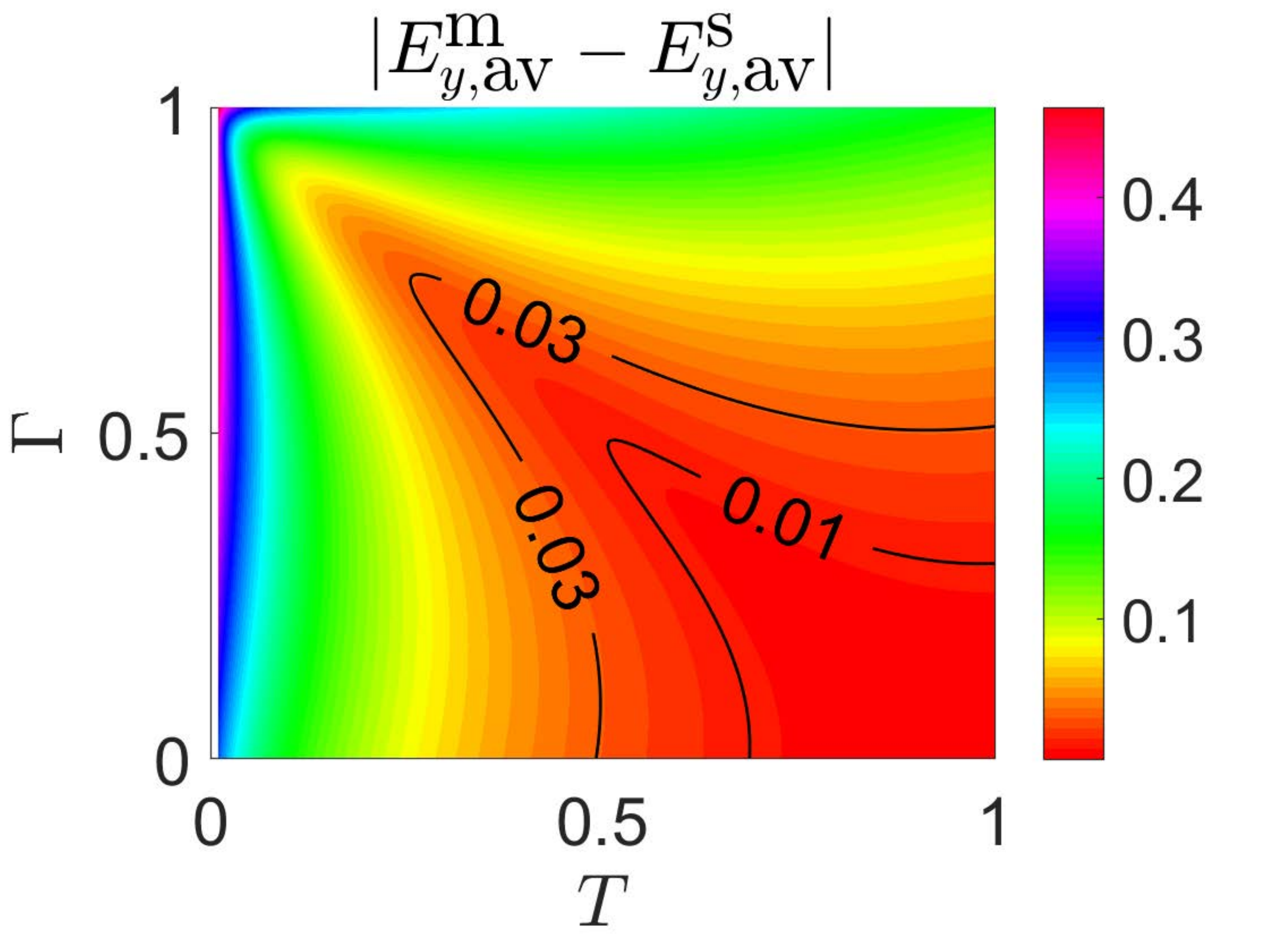}
		\caption{}\label{Fig:Diff_E}
	\end{subfigure}
	\begin{subfigure}{0.48\columnwidth}
		\centering
		\includegraphics[width=1\columnwidth]{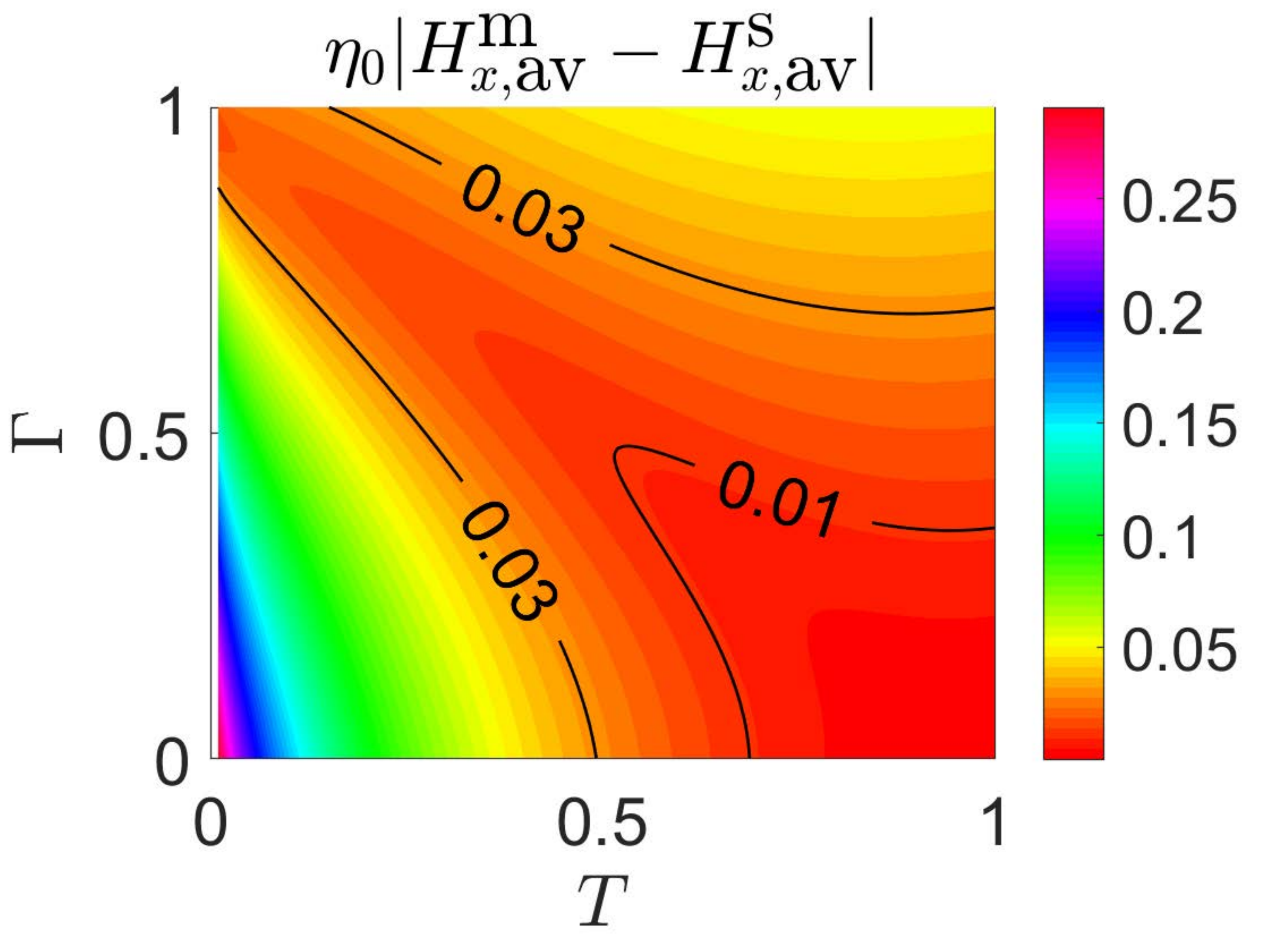}
		\caption{}\label{Fig:Diff_H}
	\end{subfigure}
	\caption{Absolute value of (a) $E_{y,\text{av}}^\text{m}/E_{y,\text{av}}^\text{s}$, (b) $H_{x,\text{av}}^\text{m}/H_{x,\text{av}}^\text{s}$, (c) $E_{y,\text{av}}^\text{m}-E_{y,\text{av}}^\text{s}$, and (d) $\eta_0(H_{x,\text{av}}^\text{m}-H_{x,\text{av}}^\text{s})$ versus $T$ and $\Gamma$ values.} \label{Ratio_difference_plots}	
\end{figure}

Figure~\ref{Ratio_difference_plots} plots the absolute value of $E_{y,\text{av}}^\text{m}/E_{y,\text{av}}^\text{s}$, $H_{x,\text{av}}^\text{m}/H_{x,\text{av}}^\text{s}$, $E_{y,\text{av}}^\text{m}-E_{y,\text{av}}^\text{s}$ and $\eta_0(H_{x,\text{av}}^\text{m}-H_{x,\text{av}}^\text{s})$ versus ($T$,$\Gamma$). The electric and magnetic field ratios and difference have similar variation trends in terms of $(T,\Gamma)$, except for $(T,\Gamma)\rightarrow (0,1)$ (PMC limit), where the two fields decouple with 	$(|E_{y,\text{av}}^\text{m}/E_{y,\text{av}}^\text{s}|,|H_{x,\text{av}}^\text{m}/H_{x,\text{av}}^\text{s}|)\rightarrow (\infty,0)$, according to Appendix~B. At the point $(T,\Gamma)=(0.5,0)$, indicated by a circle in Fig.~\ref{Fig:reflectionless_mt_parametric}, we have $|E_{y,\text{av}}^\text{m}/E_{y,\text{av}}^\text{s}| \approx |H_{x,\text{av}}^\text{m}/H_{x,\text{av}}^\text{s}| \approx 1.05$ and $|E_{y,\text{av}}^\text{m}-E_{y,\text{av}}^\text{s}| \approx |\eta_0(H_{x,\text{av}}^\text{m}-H_{x,\text{av}}^\text{s})| \approx 0.03$. The AFA [Eq.~\eqref{eq:solution}] is naturally expected to provide a satisfactory approximation in the ($T$,$\Gamma$) range where the average metasurface and slab fields are close to each other, which is increasingly the case as the metasurface effect decreases, i.e., $(T,\Gamma)\rightarrow(1,0)$.

\begin{figure}[!ht]
	\centering
	\begin{subfigure}{0.48\columnwidth}
		\centering
		\includegraphics[width=1\columnwidth]{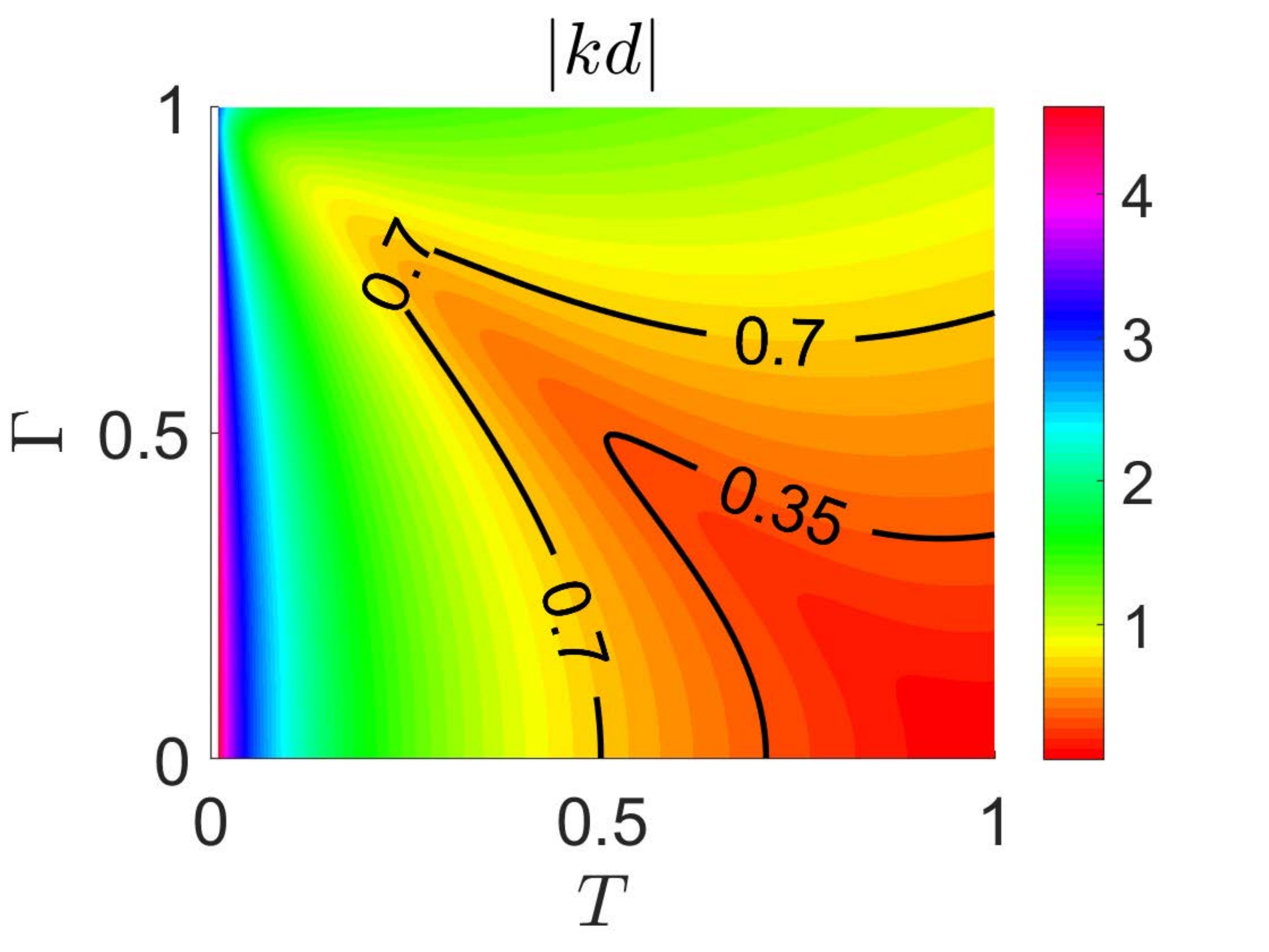}
		\caption{}\label{Fig:RT_kd_2d_plot}
	\end{subfigure}
	\begin{subfigure}{0.48\columnwidth}
		\centering
		\includegraphics[width=1\columnwidth]{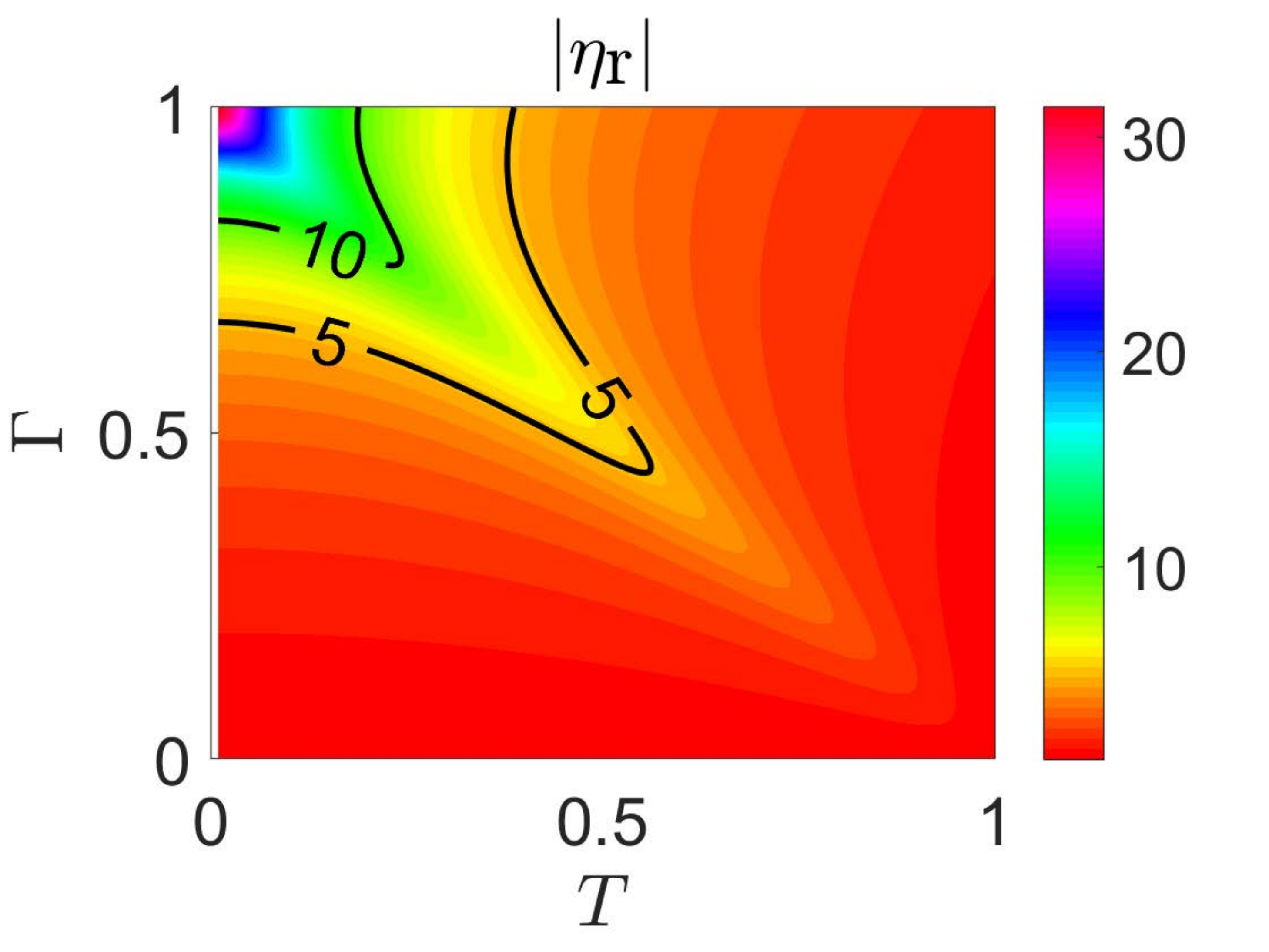}
		\caption{}\label{Fig:etar_2d_plot}
	\end{subfigure}
	\caption{Absolute value of (a)~$kd$ and (b) $\eta_\text{r}$ versus $(T,\Gamma)$.} \label{fig:kd_etar_plots}
\end{figure}

Figure~\ref{fig:kd_etar_plots} plots $|kd|$ and $|\eta_\text{r}|$ versus ($T$,$\Gamma$). According to Fig.~\ref{Fig:RT_kd_2d_plot}, for $\Gamma=0$ and $kd< 0.7\approx 2\sqrt{3}/5$, we have $T>0.5$, which is within the AFA range of validity according to Fig.~\ref{Ratio_difference_plots}, consistently with the approximation~\eqref{eq:approximated_kd}.

\begin{figure}[!ht]
	\centering
	\begin{subfigure}{0.48\columnwidth}
		\centering
		\includegraphics[width=1\columnwidth]{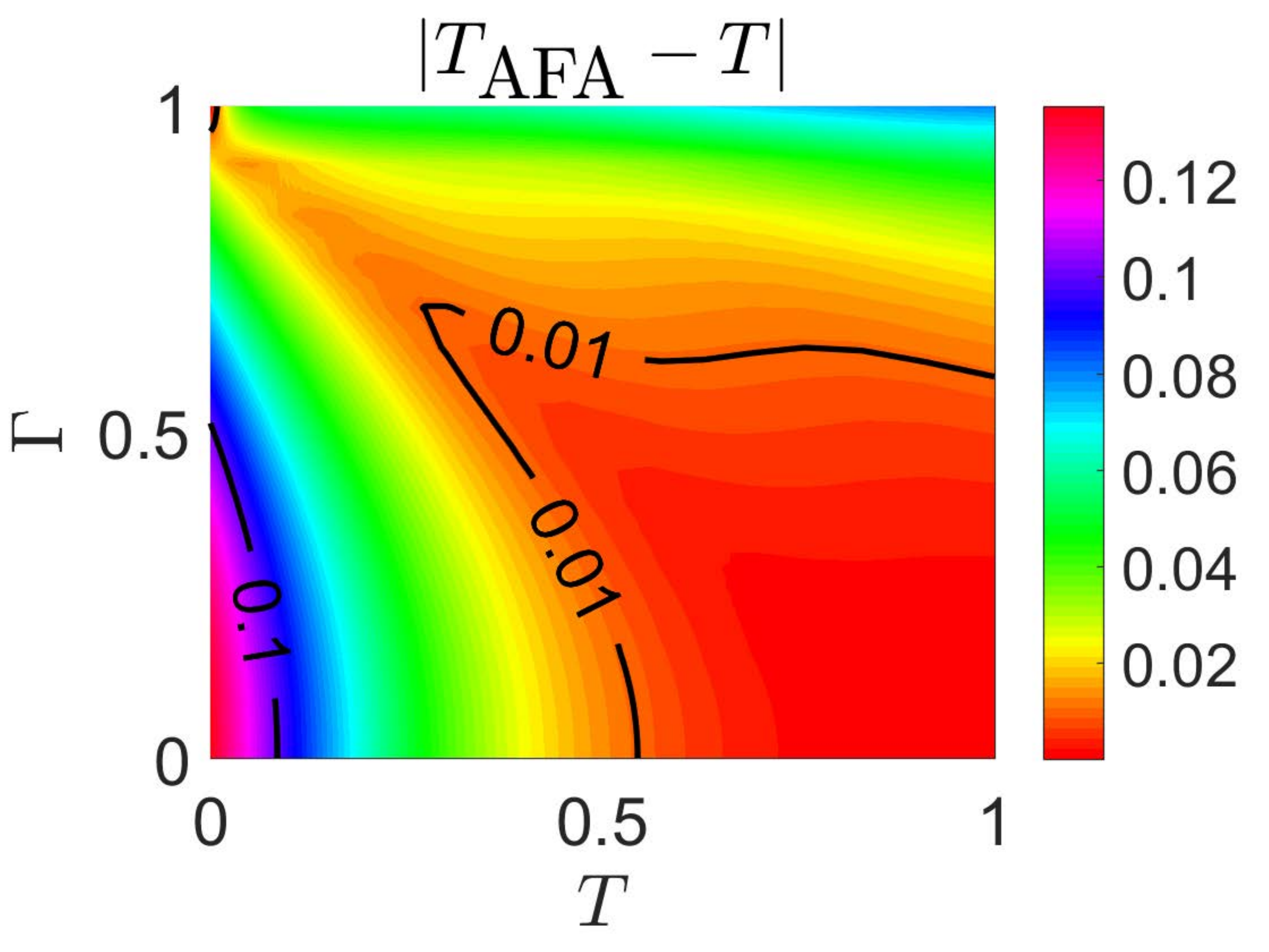}{}
		\caption{}\label{FIG:T_AFA_Exact}
	\end{subfigure}
	\begin{subfigure}{0.48\columnwidth}
		\centering
		\includegraphics[width=1\columnwidth]{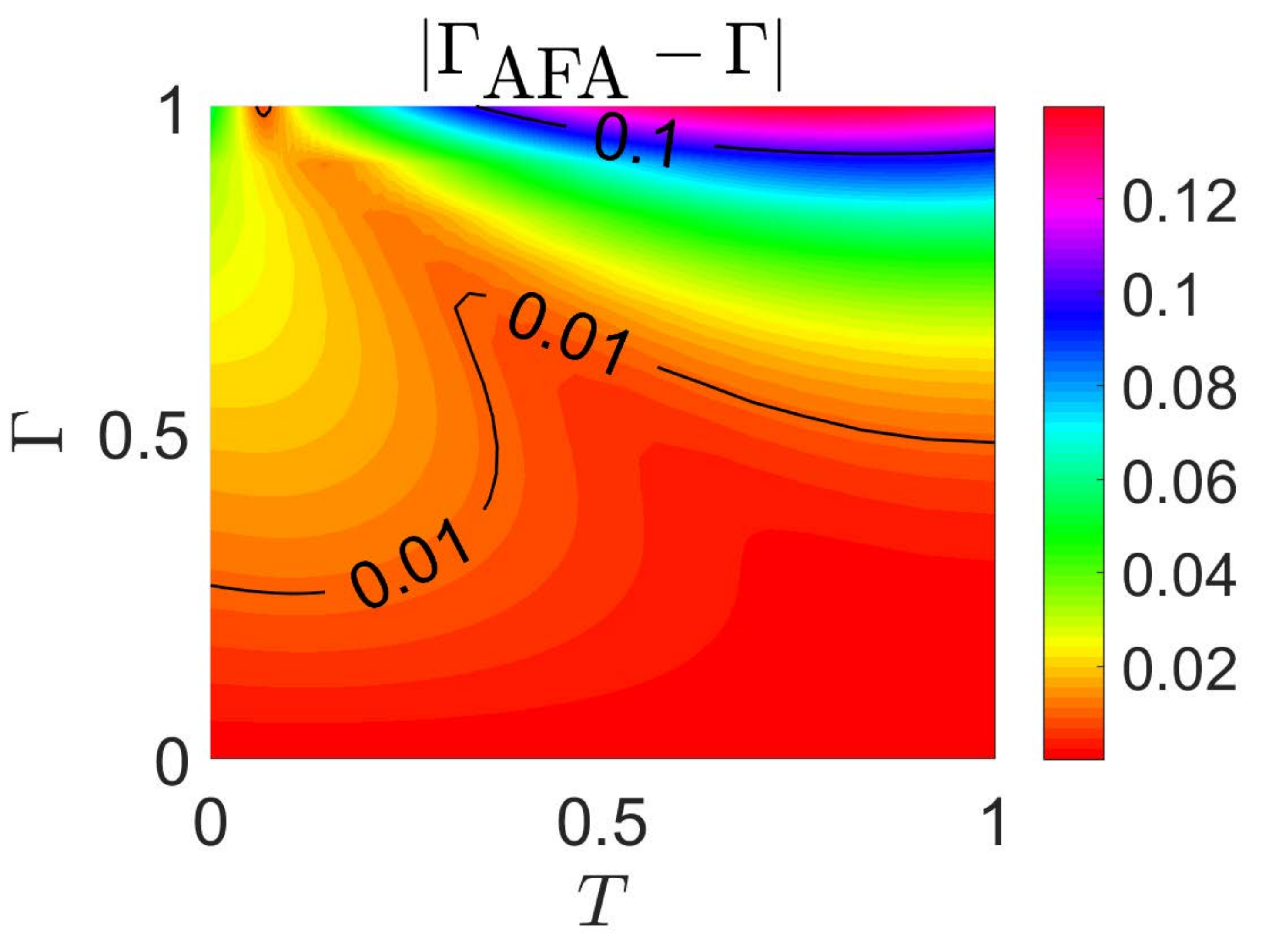}{}
		\caption{}\label{Fig:Gamma_AFA_exact}
	\end{subfigure}
	\caption{Range of validity of the AFA formula~\eqref{eq:solution} in terms of (a)~transmission and (b)~reflection coefficient differences.} \label{AFA_Exact}
\end{figure}

Finally, Fig.~\ref{AFA_Exact} plots the difference between the AFA and exact scattering parameters versus the latter, showing the $(T,\Gamma)$ range of validity of the AFA formula~\eqref{eq:solution}. The differences are found to be less than about $0.01$ in the region where $T>0.5$ and $\Gamma<0.5$. At the point ($T=0.5$, $\Gamma=0$), $|T_\text{AFA}-T| \approx 0.135$. The deviation of the AFA from the exact result quickly raises for $(T,\Gamma)\rightarrow(0,0)$, as exemplified in the matched absorber (Fig.~\eqref{FIG:ABSabsorber_COMSOL}), and $(T,\Gamma)\rightarrow(1,1)$, which would correspond to an active metasurface of gain $2$.

\section{Conclusion}\label{sec:concl}

We have investigated the possibility to model a metasurface by a thin slab of uniform permittivity and permeability. For the particular case of a uniform homoisotropic metasurface under normal plane wave incidence, we have derived an exact relation between the metasurface surface susceptibilities and the slab voluminal susceptibilities and established the related range of validity of the handy and insightful Average Field Approximation (AFA) formula in terms of the scattering parameters.

Generalizing this modeling to more complex situations, where the illumination angle could be oblique, and where the metasurface could be nonuniform and bianisotropic, would require much more complicated developments, generally involving the resolution of coupled nonlinear equations. In such situations, the complexity and non-analycity of the equivalence formulas would make the slab modeling essentially unpractical, and require the resort to metasurface-specific GSTC treatment~\cite{Yousef_FDFD_2016, Yousef_FDTD_2018, Hosseini_PLRC_FDTD_MS, Yousef_Comp_analy_ms_2018,Chamanara_TAP_04_2019}.
%


\appendices
\numberwithin{figure}{section}
\numberwithin{equation}{section}


\section{Derivation of the Exact Slab Average Fields}\label{sec:ave_fields}
We need to calculate ${E_{y,\text{av}}^\text{s}}$ (${H_{x,\text{av}}^\text{s}}$) to find the $\chi_\text{v,ee}$ ($\chi_\text{v,mm}$) in Eqs.~\eqref{eq:av_ratio_chi}. Assuming plane-wave incidence with unit electric field amplitude, the exact field \emph{inside} the slab is given by~\cite{jinau2000electromagnetic}
\begin{subequations}\label{eq:inside_slab_field}
	\begin{equation}\label{eq:inside_slab_field_gen}
	E_y^\text{s}(z)=A\left[\text{e}^{-jkz}+R\text{e}^{jk(z-d)}\right],
	\end{equation}
	where
	\begin{equation} \label{eq:Aform}
	A=\frac{2\eta_\text{r}\text{e}^{-j\left(k-k_0\right)d/2}}{\left(1+\eta_\text{r}\right)\left( 1-R^2\text{e}^{-2jkd}\right)}.
	\end{equation}
\end{subequations}

Substituting the field expression~\eqref{eq:inside_slab_field} into~\eqref{eq:Eyavs} yields
\begin{equation}\label{eq:step1_substition}
E_{y,\text{av}}^\text{s}=\frac{A}{d}\int\limits_{-d/2}^{d/2}\left[e^{-jkz}+\frac{1-\eta_\text{r}}{1+\eta_\text{r}} e^{jk(z-d)}\right]\mathrm{d}z,
\end{equation}
which can be analytically calculated as
\begin{equation}\label{eq:integ_field_slab}
E_{y,\text{av}}^\text{s}=\frac{4A\sin\left( \frac{kd}{2}\right)}{kd\left(1+\eta_\text{r}\right)}\left[ \cos\left(\frac{kd}{2}\right)+j\eta_\text{r}\sin\left(\frac{kd}{2}\right) \right]\text{e}^{-j\frac{kd}{2}},
\end{equation}
and simplifies to 

\begin{subequations}\label{eq:step2_simplification}
	\begin{equation}\label{eq:Eyavs_evaluated}
	{E_{y,\text{av}}^\text{s}}=\frac{2A\sin\left(\frac{kd}{2}\right)}{kd}\left(1+\frac{1-\eta_\text{r}}{1+\eta_\text{r}}e^{-jkd}\right).
	\end{equation}
	Similarly, the average magnetic field is given by
	\begin{equation}\label{eq:Hxavs_evaluated}
	{H_{x,\text{av}}^\text{s}}=\frac{-2A\sin\left(\frac{kd}{2}\right)}{kd\eta}\left(1-\frac{1-\eta_\text{r}}{1+\eta_\text{r}}e^{-jkd}\right),
	\end{equation}
\end{subequations}
where $\eta=\eta_\text{r}\eta_0$ is the wave impedance in the slab medium.

\section{Average Fields for ($T$,$\Gamma$)$\rightarrow$(0,1)}\label{sec:approx_ave_fields}

For $\Gamma=1$ and $T\approx 0$, we have
\begin{equation}\label{eq:m_ave_fields}
(E_{y,\text{av}}^\text{m},H_{x,\text{av}}^\text{m}) \approx (1,-T/(2\eta_0))
\end{equation}
and, according to~\eqref{eq:step2_simplification}, 
\begin{equation}\label{eq:s_ave_fields}
(E_{y,\text{av}}^\text{s},H_{x,\text{av}}^\text{s}) \approx \frac{2}{jkd} (1,-1/\eta).
\end{equation}  

Dividing Eq.~\eqref{eq:m_ave_fields} by Eq.~\eqref{eq:s_ave_fields} and then taking the absolute value yields then
\begin{equation}\label{eq:fields_ratios}
\left(\left|\frac{E_{y,\text{av}}^\text{m}}{E_{y,\text{av}}^\text{s}}\right|,\left|\frac{H_{x,\text{av}}^\text{m}}{H_{x,\text{av}}^\text{s}}\right|\right) \approx \frac{|kd|}{2}\left(1,\frac{T|\eta_\text{r}|}{2}\right).
\end{equation}

According to~\eqref{eq:R}, at $\Gamma=1$ and $T \approx 0$, we have $R \approx -1+jk_0d$, and therefore, from~\eqref{eq:RR}, $\eta_\text{r}\approx -j{2}/{k_0d}$. Moreover, at $\Gamma=1$ and $T \approx 0$, we also find from~\eqref{eq:exp_kd} that $kd \rightarrow -j\infty$, so that, according to~\eqref{eq:slab_trans_ref}, we have $T \approx \text{e}^{-jkd}$. As a result, Eq.~\eqref{eq:fields_ratios} leads to   
\begin{equation}\label{eq:fields_ratios_reduced1}
\left(\left|\frac{E_{y,\text{av}}^\text{m}}{E_{y,\text{av}}^\text{s}}\right|,\left|\frac{H_{x,\text{av}}^\text{m}}{H_{x,\text{av}}^\text{s}}\right|\right) \approx \frac{|kd|}{2} \left(1,\frac{\text{e}^{-|kd|}}{k_0d}\right).
\end{equation}

As $T\rightarrow{0}$, $|kd|\rightarrow\infty$, so that Eq.~\eqref{eq:fields_ratios_reduced1} reduces to    
\begin{equation}\label{eq:fields_ratios_reduced2}
\left(\left|\frac{E_{y,\text{av}}^\text{m}}{E_{y,\text{av}}^\text{s}}\right|,\left|\frac{H_{x,\text{av}}^\text{m}}{H_{x,\text{av}}^\text{s}}\right|\right) \rightarrow (\infty,0).
\end{equation}

\bibliographystyle{IEEEtran}
\bibliography{MS_MODELING_THIN_SLABS}

\end{document}